\documentclass[prd,showpacs,preprint,eqsecnum,amsmath,amssymb,nofootinbib]{revtex4-1}

\usepackage{hyperref}
\usepackage{amsfonts}
\usepackage{mathrsfs}
\usepackage{amssymb}
\usepackage{graphicx, epsfig, amsmath}
\usepackage{dcolumn}
\usepackage{bm}

\newcommand{\la}{\langle}
\newcommand{\ra}{\rangle}
\newcommand{\w}{\omega}

\newcommand{\be}{\begin{equation}}
\newcommand{\ee}{\end{equation}}
\newcommand{\bea}{\begin{eqnarray}}
\newcommand{\eea}{\end{eqnarray}}
\newcommand{\bes}{\begin{subequations}}
\newcommand{\ees}{\end{subequations}}

\begin{document}
\title{Method to compute the stress-energy tensor for a quantized scalar field when a black hole forms from the collapse of a null shell}
\author{Paul~R.~Anderson${}^{1}$}
\email{anderson@wfu.edu}

\author{Shohreh Gholizadeh Siahmazgi${}^{1}$}
\email{ghols18@wfu.edu}

\author{Raymond D.\ Clark${}^{1}$}

\author{Alessandro~Fabbri${}^{2, 3}$}
\email{afabbri@ific.uv.es}

\affiliation{${}^1$Department of Physics, Wake Forest University, Winston-Salem, North Carolina 27109, USA}
\affiliation{${}^2$Departamento de F\'isica Te\'orica and IFIC, Universidad de Valencia-CSIC, C. Dr. Moliner 50, 46100 Burjassot, Spain}
\affiliation{${}^3$Universit\'e Paris-Saclay, CNRS/IN2P3, IJC Lab, 91405 Orsay Cedex, France}

\date{\today}

\begin{abstract}

A method is given to compute the stress-energy tensor for a massless minimally coupled scalar field in a spacetime where a black hole
forms from the collapse of a spherically symmetric null shell in four dimensions.  Part of the method involves matching the modes for the {\it in} vacuum state
to a complete set of modes in Schwarzschild spacetime.  The other part involves subtracting from the unrenormalized expression for the stress-energy
tensor when the field is in the {\it in} vacuum state, the corresponding expression when the field is in the Unruh state and adding to this the renormalized
stress-energy tensor for the field in the Unruh state.  The method is shown to work in the two-dimensional case where the results are known.

\end{abstract}

\maketitle

\section{Introduction}

The stress-energy tensor of a quantized field is an extremely useful tool for studying quantum effects in curved space because it takes both particle
production and vacuum polarization into account.  It can be computed in a background spacetime to obtain the energy density, pressure, etc.
for a quantum field in that spacetime.  It can also be used in the context of semiclassical gravity to compute the backreaction of the quantum field on the
spacetime geometry.

For black holes in four-dimensional, 4D, spacetimes, the full stress-energy tensor must be computed numerically.  This is a difficult task that has to date only been done without other approximations for the cases of static spherically symmetric black holes~\cite{fawcett,howard-candelas,howard,jensen-ottewill,jmo,jensen-et-al,ahs1,ahs2,ahl,choag,abf,breen-ottewill,levi-ori,levi,Zilberman-Levi-Ori} and the stationary Kerr metric~\cite{duffy-ottewill,levi-et-al-kerr}.
However, because of the difficulty involved, to our knowledge, no one has numerically computed the full stress-energy tensor for a quantized field in a 4D spacetime in which a black hole forms from collapse.  This is important because there can be a significant difference between the stress-energy tensor for a quantum field in a 2D versus a 4D spacetime such as that found for a massless minimally coupled scalar field in an extreme Reissner-Nordstrom spacetime~\cite{trivedi,ahl}.

In this paper we present a  method to compute the renormalized stress-energy tensor, $\la {\rm in}| T_{ab} | {\rm in} \ra$, for a massless minimally coupled scalar field in the case that
a black hole forms from the collapse of a spherically symmetric null shell.  The method works in the region outside both the shell and the event horizon.

In the region outside the null shell, Birkhoff's theorem ensures that the metric is that for Schwarzschild spacetime~\eqref{metric-sch}.
In the region inside the shell the space is flat.  Thus in both regions the mode equation for the quantum field is separable and inside the shell its solutions are
known analytically.  This allows for a numerical computation of the stress-energy tensor for the field in which only ordinary differential equations need to be solved
numerically.

For the collapsing null shell model, the initial vacuum state of the quantum field is well defined and
the main complication that occurs is due to the propagation of the modes across the null shell surface.  The crux of our method involves the expansions of the {\it in}
modes in terms of a complete set of solutions to the mode equation in the region outside the shell.

The stress-energy tensor for the quantum field is obtained by expanding the quantum field in terms of a complete set of modes.
This expansion is substituted into the formula for the stress-energy tensor of the corresponding classical field and the expectation value is computed.
If the field is in the {\it in} vacuum state then the result is an expression which involves sums and integrals over the mode functions for the {\it in} state and their derivatives.  After the renormalization counterterms are subtracted off, the resulting stress-energy tensor is finite and can be computed. This
is straightforward inside the null shell since the mode functions are known analytically and for the {\it in} state, the result is that the stress-energy tensor is equal to zero.

Outside the null shell and the event horizon the {\it in} modes do not assume a simple form in 4D.  One way to extend them to this region would be to use their values on the null shell trajectory along with their values at past null infinity, $\mathscr{I}^{-}$, as initial data for direct numerical computations of these mode function using numerical methods for partial differential equations.  A second way, which is the one we adopt here, is expand each of the {\it in} modes in terms of a complete set of modes in Schwarzschild spacetime.  The complete set of modes can be obtained by solving ordinary differential equations since the mode equation is separable.  The radial parts of these modes and some of the matching parameters must be computed numerically.  The mode matching has been tested in the 2D case where there is no effective potential in the mode equation.  It has also been partially tested for spherically symmetric modes in 4D when the effective potential is modeled as a delta function and for the full effective potential.

When the expansions for the {\it in} modes are substituted into the formula for the unrenormalized stress-energy tensor one finds
a triple integral over the mode functions which form a complete set in Schwarzschild spacetime.  Renormalization can be accomplished by subtracting the corresponding
expression that occurs for the Unruh state, then adding that expression back and subtracting the renormalization counterterms.
  The result is the sum of two finite tensors, the difference between the stress-energy tensor for the {\it in} state and the Unruh state and the renormalized stress-energy tensor for the Unruh state.  The latter has been numerically computed for the masslesss minimally coupled scalar field in~\cite{levi-ori,levi}.
This type of renormalization scheme has been used to compute the stress-energy tensors in Schwarzschild spacetime in the Unruh state for the conformally coupled massless scalar field~\cite{elster,jmo} and for the massless spin $1$ field~\cite{jmo}.

We have tested our method by numerically computing the stress-energy tensor in the collapsing null shell spacetime in 2D where the answer can be compared with
previous analytic calculations for the Unruh state~\cite{Davies-Fulling-Unruh} and the {\it in} vacuum state for the collapsing null shell spacetime~\cite{hiscock, Fabbri:2005mw}.
Our results are in agreement with those calculations.

In Sec. II we introduce the collapsing null shell model and then discuss the modes for a massless minimally coupled scalar field in both the null shell spacetime and pure Schwarzschild spacetime.  A detailed description of the method, including the computation of the $in$ modes and the  renormalization of the stress-energy tensor is given in Sec. III.  General expressions for the matching coefficients in the 4D case are derived in Sec. IV followed by examples where the matching method is tested.
Formulas needed for the computation of the stress-energy tensor in the 4D case are derived in the first part of Sec. V.  In the second part the stress-energy tensor is computed using our method in the 2D case and compared with previous calculations.
Sec. VI contains a summary of our results.  The appendices contain some details of a proof and some derivations that are used in the 2D examples in Secs. IV and V.
Throughout the paper, we use the sign conventions of~\cite{Misner} and units are chosen such that $\hbar=c=G=1$.

\section{Collapsing Null Shell Model}

The type of quantum field we consider is a massless minimally coupled scalar field.  In a static spherically symmetric spacetime it can be expanded in terms of a complete set of modes $f_{\w \ell m}$ such that
\be \phi = \sum_{\ell = 0}^\infty  \sum_{m = -\ell}^\ell \int_0^\infty [a_{\w \ell m} f_{\w \ell m} + a^\dagger_{\w \ell m} f^{*}_{\w \ell m}] \;, \label{Phi-expansion} \ee
with $a_{\w \ell m}$ an annihilation operator.  The modes are solutions to the equation
\be \Box f_{\w \ell m} = 0  \;, \label{Box-f} \ee
and have the general form
\be f_{\w \ell m} = N \frac{Y_{\ell, m}(\theta, \phi)}{r } \psi(\tau,r) \;. \label{f-def-1} \ee
The normalization constant $N$ is fixed by the condition
\be  (f_{\w \ell m}, f_{\w' \ell' m'}) = \delta_{\ell,\ell'} \delta_{m,m'} \delta(\w-\w') \;. \label{f-norm} \ee
The scalar product is defined by
\bea
  (f_1, f_2) = -i\int_{\Sigma} d \Sigma \, n^\mu [f_1(x)\overset{\leftrightarrow}{\partial_\mu}f_2^{*}(x)]
 \;, \label{scalar-products}
\eea
where $n^{\mu}$ is a future-directed unit vector orthogonal to the spacelike (or null) hypersurface $\Sigma$ and $d\Sigma$ is the volume element in $\Sigma$. The hypersurface $\Sigma$ is taken to be a Cauchy surface and it is assumed that the spacetime is globally hyperbolic.

\begin{figure}[h]
\centering
\includegraphics [trim=0cm 20cm 0cm 0cm,clip=true,totalheight=0.3\textheight] {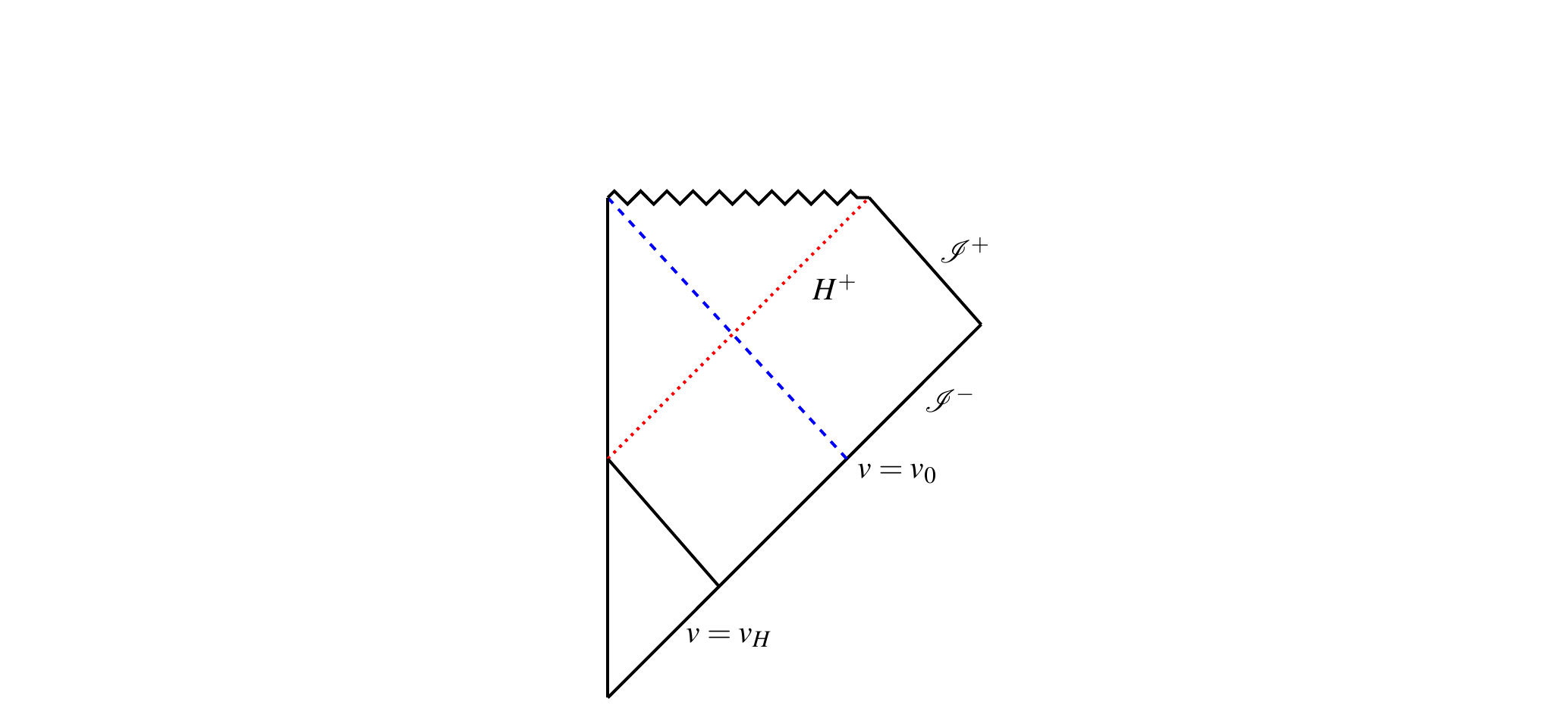}
\caption{Penrose diagram for a spacetime in which a null shell collapses to form a spherically symmetric black hole.  The trajectory of the shell (dashed blue curve) is $v = v_0$.  The horizon, $H^{+}$, is the dotted red curve.}
\label{BH_4D_penrose}
\end{figure}

We consider a model in which a spherically symmetric black hole forms from the collapse of a null shell.  The Penrose diagram is shown in Fig.~\ref{BH_4D_penrose}.  The metric inside the shell is the flat space metric
\be
ds^2 = -dt^2 + dr^2 + r^2 d \Omega^2    \;, \label{metric-flat}
\ee
and the metric outside the shell is the Schwarzschild metric
\be
ds^2 = -\left(1-\frac{2M}{r} \right) dt_s^2
+ \left( 1-\frac{2M}{r} \right)^{-1} dr^2 + r^2 d \Omega^2  \;. \label{metric-sch}
\ee
It is useful inside the shell to define the radial null coordinates
\bes \bea
u &=& t - r \;, \label{u-flat}\\
 v &=& t + r  \;,\, \label{v-flat} \eea
\label{u-v-flat}
\ees
and outside the shell to define the radial null coordinates
\bes \bea
u_s &=& t_s - r_* \;, \label{u-sch-def}\\
  v &= & t_s + r_* \;, \label{v-sch-def}
 \eea   \label{u-v-sch}
\ees
where
\bea
  r_* &= & r + 2M \log \left(\frac{r-2M}{2M} \right)  \;.
\label{rstar-def}
\eea
The angular coordinates, the $r$ coordinate and the $v$ coordinate are continuous across the shell.  The time coordinate is discontinuous as are $u$ and $u_s$ .
The relationship between the latter two is~\cite{m-p,Fabbri:2005mw}
\be
u_s = u - 4M \log \left( \frac{v_H-u}{4 M} \right)  \;,
\label{us-u}
\ee
with
\be
v_H \equiv v_0 - 4M  \;.
\label{vH-def}
\ee
Note that the value of $u$ on the event horizon is $v_H$.  Inverting, one finds that~\cite{mirror-bh}
\be
u = v_H - 4 M \, W\left[\exp\left(\frac{v_H - u_s}{4 M}\right) \right]  \;,
\label{u-us}
\ee
with $W$ the Lambert W function.

For all of the mode functions that we consider the normalization constant is
\be N = \frac{1}{\sqrt{4 \pi \w}} \;. \label{Normalization-constant} \ee
Inside the shell, using~\eqref{f-def-1} and the metric~\eqref{metric-flat} in~\eqref{Box-f} one finds mode functions of the form
\bes \bea f_{\w \ell m} &=&  \frac{Y_{\ell m}(\theta,\phi)}{r \sqrt{4 \pi \w}} \psi_{\w \ell}(t,r)\;, \label{f-form-in} \\  \nonumber \\
    \psi_{\w \ell} &=& e^{-i \w t} \chi_{\w \ell}(r) \;,  \\ \nonumber \\
     \frac{d^2 \chi_{\w \ell}}{d r^2} &=& - \left[\w^2 - \frac{\ell(\ell+1)}{r^2} \right] \chi_{\w \ell} \;. \label{mode-eq-inside}
\eea \ees
Outside the shell, using~\eqref{f-def-1}, the metric~\eqref{metric-sch}, and~\eqref{rstar-def} in~\eqref{Box-f} one finds mode functions of the form
\bes \bea f_{\w \ell m} &=&  \frac{Y_{\ell m}(\theta,\phi)}{r \sqrt{4 \pi \w} } \psi_{\w \ell}(t_s,r)\;,  \label{f-form-out}  \\ \nonumber \\
    \psi_{\w \ell} &=& e^{-i \w t_s} \chi_{\w \ell}(r) \;, \label{psi-outside} \\ \nonumber \\
     \frac{d^2 \chi_{\w \ell}}{d r_{*}^2} &=& -  \left[\w^2 - \left(1 - \frac{2M}{r} \right) \left( \frac{2 M}{r^3}+  \frac{\ell(\ell+1)}{r^2} \right) \right] \chi_{\w \ell} \;.\label{chi-eq-outside}  \eea \label{modes-outside} \ees
Note that in pure Schwarzschild spacetime outside of the past and future horizons, the form of the modes and the equation they satisfy are also given by~\eqref{modes-outside}.

The {\it in} vacuum state in the null shell spacetime is specified by
\be \psi^{\rm in}_{\w \ell} = e^{-i \w v} \; \label{in-vacuum}\ee
on past null infinity $\mathscr{I^{-}}$ and by requiring that $\psi_{\w \ell}$ vanish at $r = 0$ in the region inside the shell so that $f_{\w \ell m}$ is regular there.
The solution that has these properties is
\be \psi^{\rm in}_{\w \ell} =  C_{\ell} \, e^{-i \w t}\, \w r  j_\ell(\w r) \;, \label{psi-in-a} \ee
where $C_{\ell}$ is a normalization constant and $j_\ell $ is a spherical Bessel function.  The condition~\eqref{in-vacuum} fixes the value of $C_{ \ell}$.  For example, for $\ell = 0$, it is easy to show that $C_{0} = -2 i $ and
\be \psi^{\rm in}_{\w 0} = e^{-i \w v} - e^{-i \w u} \;. \label{psi-in-0} \ee

\section{Method to compute the stress-energy tensor}
\label{sec:method-Tab}

The stress-energy tensor for the quantized massless minimally coupled scalar field, $\la T_{ab} \ra$, is to be computed for the {\it in} vacuum state in the region outside the null shell and outside the event horizon. The stress-energy tensor for the classical field is
\be T_{ab} = \partial_a \Phi  \partial_b \Phi - \frac{1}{2} g_{ab} g^{cd} \partial_c \Phi  \partial_d \Phi  \;. \label{Tab-class} \ee
To compute $\la {\rm in}| T_{ab} |{\rm in} \ra$, one can substitute~\eqref{Phi-expansion} into~\eqref{Tab-class}, use the complete set of modes for the {\it in} vacuum state $f^{\rm in}_{\w \ell m}$, and compute the expectation value.  There are two things which make this difficult. One is computing the modes $f^{\rm in}_{\w \ell m}$ in the region outside the shell and the other is renormalizing the stress-energy tensor.  Our method to compute the stress-energy tensor provides one way to overcome these difficulties.

First, we renormalize by subtracting from the unrenormalized expression for the stress-energy tensor for the {\it in} vacuum state, the unrenormalized stress-energy tensor for the Unruh state.  Since the renormalization counterterms are local and thus do not depend on the state of the quantum field, this quantity will be finite.  Then we add back the
 unrenormalized stress-energy tensor for the Unruh state and then subtract from it the renormalization counter terms.
 Schematically one can write
 \bea \la {\rm in}| T_{ab}| {\rm in} \ra_{\rm ren} &=& \Delta \la T_{ab} \ra +   \la U| T_{ab} | U \ra_{\rm ren} \;,  \nonumber \\
               \Delta \la T_{ab} \ra  &=&  \la {\rm in}| T_{ab} |{\rm in} \ra_{\rm unren} - \la U| T_{ab} | U \ra_{\rm unren} \;. \label{Tab-Unruh-sub} \eea
The quantity $\la U| T_{ab} | U \ra_{\rm ren}$ has been numerically computed for a massless minimally coupled scalar field in Schwarzschild spacetime~\cite{levi-ori, levi}.
Thus what remains is to compute the difference between the unrenormalized expressions.  To do that it is necessary to discuss the computation of the mode functions for the quantum field that are relevant for the {\it in} and Unruh states.
It is worth pointing out that the computation of $\la U| T_{ab} | U \ra_{\rm unren}$ done in~\cite{levi-ori,levi} was done for pure Schwarzschild spacetime outside the event horizon.  However the computation we wish to do for
$\la {\rm in}| T_{ab}| {\rm in} \ra_{\rm ren}$ is for the null shell spacetime outside both the shell and the horizon.  The reason that there is no problem is that the renormalization counterterms are local and so are the same in this part of the null shell spacetime as they are in pure Schwarzschild spacetime.

Analytic expressions for the mode functions in the {\it in} vacuum state, $f^{\rm in}_{\w \ell m}$
inside the shell are given in~\eqref{psi-in-a}.  However, it is not easy to continue these to the region outside the shell because the time coordinate $t$ and the right moving radial null coordinate $u$ are not continuous across the shell.  As a result, the modes $f^{\rm in}_{\w \ell m}$, while still being solutions to~\eqref{Box-f}, do not have the form~\eqref{f-form-out} outside the shell. However, the known solutions inside the null shell along with their behavior on $\mathscr{I}^{-}$ can be used to fix the initial data on a Cauchy surface in the null shell spacetime.  The Cauchy surface we consider here, consists of the part of $\mathscr{I}^{-}$ with $v_0 \le v < \infty$ along with the trajectory of the null shell.  This initial data could be used for a numerical calculation of the partial differential equation satisfied by $f^{\rm in}_{\w \ell m}$ outside the shell.  Alternatively, one can expand  $f^{\rm in}_{\w \ell m}$ in terms of a complete set of modes in the region outside the shell and use the data on the Cauchy surface to determine the matching coefficients.

\begin{figure}[h]
\centering
\includegraphics   [trim=0cm 0cm 0cm 0cm,clip=true,totalheight=0.25\textheight]{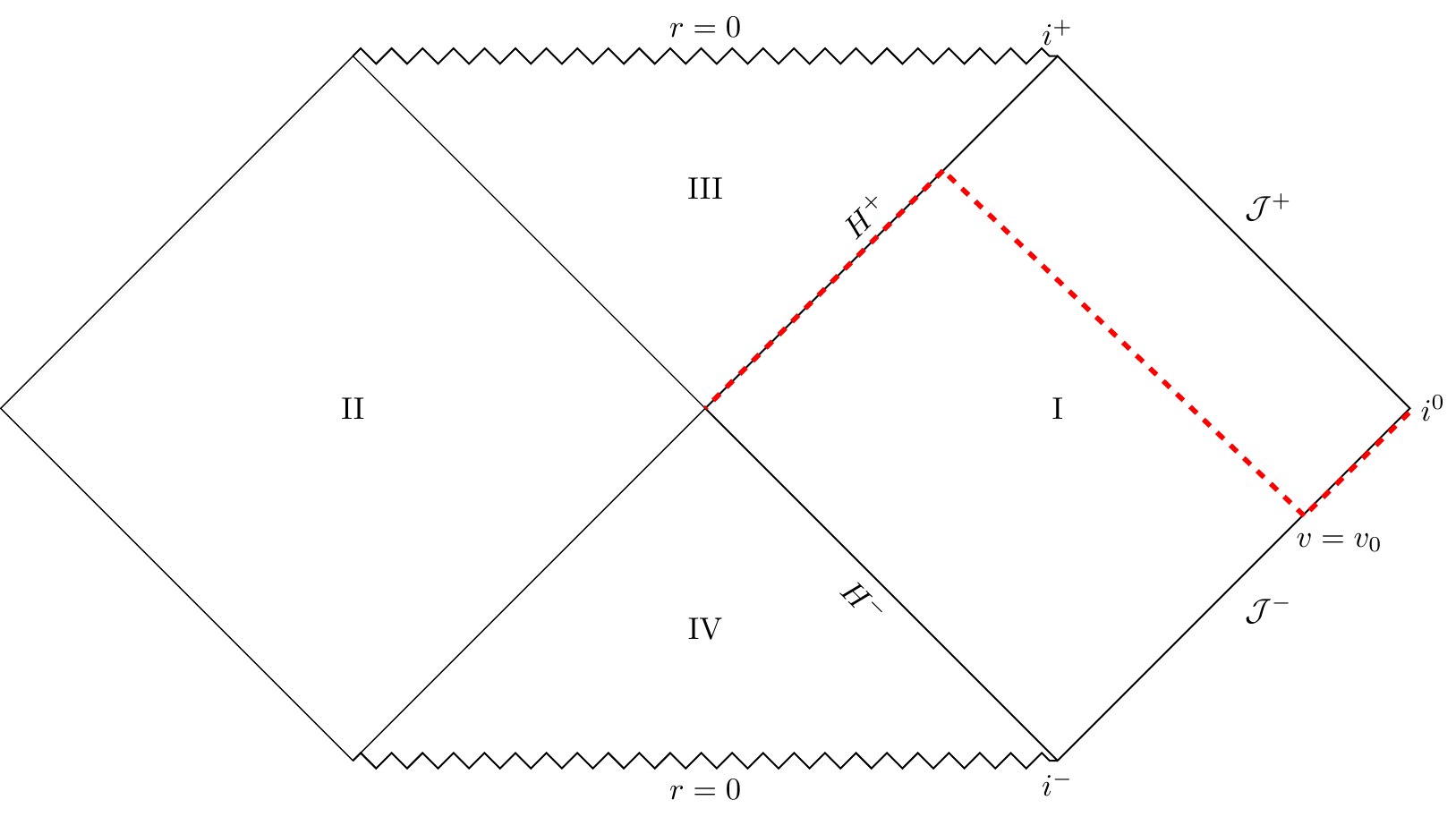}
\caption{Penrose diagram for Schwarzschild spacetime showing the Cauchy surface used for matching the {\it in} modes in the null shell spacetime
to a complete set of modes in Schwarzschild spacetime in the region outside the past and future horizons.  The Cauchy surface is denoted by the dashed red curve. }
\label{fig:Cauchy}
\end{figure}

Here we take a variation of the latter approach by noting that the spacetime geometry outside the shell is the Schwarzschild geometry.  Because of this, it is possible to do the matching in the corresponding part of Schwarzschild spacetime.  The advantage of this is that the matching can be to a complete set of modes in the region outside the horizon in Schwarzschild spacetime.  These modes are well understood and straight-forward to work with numerically.  The disadvantage is that the relevant part of the Cauchy surface in the null shell spacetime discussed above does not form a Cauchy surface in the Schwarzschild spacetime.  This can be remedied by adding a segment along the future horizon with $-\infty < v \le v_0$.  The result is a Cauchy surface for the part of Schwarzschild spacetime that is outside of the past and future horizons.  It is illustrated in Fig.~\ref{fig:Cauchy}.  It is worth noting that the part of the Cauchy surface on the future horizon is not causally connected with the region outside the future horizon and outside the surface $v = v_0$.  The corresponding region in the null shell spacetime is the region where we want to compute the stress-energy tensor.  Thus any initial data
can be used for the mode function $f^{\rm in}_{\w \ell m}$ on that surface so long as $f^{\rm in}_{\w \ell m}$ is continuous at the point where the future horizon intersects the part of the Cauchy surface with $v = v_0$.

For the matching we find it most convenient to choose the complete set of modes that consists of modes that are positive frequency on the future horizon $H^{+}$ (labeled by $f^{H^{+}}_{\w \ell m}$)  along with modes that are positive frequency on future null infinity $\mathscr{I}^{+}$ (labeled by $f^{\mathscr{I}^{+}}_{\w \ell m}$).

The Unruh state is defined in Schwarzschild spacetime by considering modes that are positive frequency on past null infinity $\mathscr{I}^{-}$ and labeled by $f^{\mathscr{I}^{-}}_{\w \ell m}$ along with modes labeled $f^{K}_{\w \ell m}$ that are positive frequency with respect to the Kruskal null coordinate
\be U = - \frac{e^{-\kappa u}}{\kappa}   \label{U-Kruskal}  \ee
on the past horizon $H^{-}$.  Here $\kappa = (4 M)^{-1}$ is the surface gravity of the black hole and $M$ is its mass.  As shown below, the modes $f^{K}_{\w \ell m}$  can be expanded in terms of modes that are positive frequency with respect to $u$ on $H^{-}$ and that we label as $f^{H^{-}}_{\w \ell m}$.  Note that together the modes $f^{\mathscr{I}^{-}}_{\w \ell m}$ and $f^{H^{-}}_{\w \ell m}$ form a complete set of modes in the part of Schwarzschild spacetime that is outside the past and future horizons.

\subsection{Properties of mode functions in Schwarzschild spacetime}

The general form of the mode functions in Schwarzschild spacetime along with the equation satisfied by the radial mode functions is given in~\eqref{modes-outside}.  The two
complete sets of modes that we work with are defined for values of $\w$ in the range $0 \le \w < \infty$.
The radial functions for these two sets of modes can be defined in terms of two linearly independent solutions to the radial mode equation~\eqref{modes-outside} with the properties
\bes \bea \chi^\infty_R &\to& e^{i \w r_*} \qquad r_* \to \infty \;,\label{chi-infinity-R-def} \\
\chi^\infty_L &\to& e^{ -i \w r_*} \qquad r_* \to \infty \label{chi-infinity-L-def}\;. \eea \label{chiRL-def} \ees
Near the event horizon they have the behaviors~\cite{rigorous}
\bes \bea \chi^\infty_R &\to& E_R(\w) e^{i \w r_*} + F_R(\w) e^{-i \w r_*} \;, \label{chiR-hor} \\
      \chi^\infty_L &\to& E_L(\w) e^{i \w r_*} + F_L(\w) e^{-i \w r_*} \;, \label{chiL-hor} \eea \label{chiRL-hor} \ees
where $E_R$, $E_L$, $F_R$, and $F_L$ are scattering parameters that can be determined numerically.\footnote{The subscripts $r$ and $l$ in~\cite{rigorous} have been changed here to $R$ and $L$ respectively.}

As discussed above, one complete set of modes consists of modes that are positive frequency  on the past horizon and modes that are positive frequency on past null infinity.
On the past horizon $\psi_{\w \, \ell}^{H^{-}} = e^{-i \w u_s}$ and $\psi_{\w \, \ell}^{\mathscr{I}^{-}} = 0$ while on past null infinity $\psi_{\w \, \ell}^{\mathscr{I}^{-}} = e^{-i \w v}$ and $\psi_{\w \, \ell}^{H^{-}} = 0$.
 These modes  are scattered by the effective potential.  It is straight-forward to show that~\cite{rigorous}
\bes \bea \chi^{H^{-}}_{\w \ell} &=& \frac{ \chi^\infty_R}{E_R} \;, \label{chi-H-minus-def} \\
         \chi_{\w \ell}^{\mathscr{I}^{-}} &=&  \chi^\infty_L - \frac{E_L}{E_R} \chi^\infty_R  \;. \label{chi-scri-minus-def}
\eea \label{chi-H-scri-def}\ees

The other complete set of modes that we will use consists of modes that are positive frequency on the future horizon and those that are positive frequency on future null infinity.  On the future horizon $\psi_{\w \, \ell}^{H^{+}} = e^{-i \w v}$ and $\psi_{\w \, \ell}^{\mathscr{I}^{+}} = 0$ while on future null infinity $\psi_{\w \, \ell}^{H^{+}} = 0$ and $\psi_{\w \, \ell}^{\mathscr{I}^{+}} = e^{-i \w u_s}$.
Going backwards in time these modes scatter due to the effective potential.   It is straight-forward to show that
\bes \bea \chi^{H^{+}}_{\w \ell} &=& \frac{1}{F_L} \chi^\infty_L \;, \label{chi-H-plus-def}  \\
\chi^{\mathscr{I}^{+}}_{\w \ell} &=& \chi^\infty_R - \frac{F_R}{F_L} \chi^\infty_L  \;.  \label{chi-scri-plus-def} \eea \ees

Since the method used to compute the stress-energy tensor involves subtracting the unrenormalized stress-energy tensor for the Unruh state it is useful to write the modes associated with this state, $f^{H^{-}}$ and $f^{\mathscr{I}^{-}}$ in terms of $f^{H^{+}}$  and $f^{\mathscr{I}^{+}}$ .
\bes \bea \chi^{H^{-}}_{\w \ell} &=& \frac{1}{E_R} ( F_R \,  \chi^{H^{+}}_{\w \ell} +  \chi^{\mathscr{I}^{+}}_{\w \ell} )  \;, \label{chi-H-minus-2} \\
\chi_{\w \ell}^{\mathscr{I}^{-}} &=& \frac{1}{E_R} (\chi^{H^{+}}_{\w \ell} - E_L \, \chi^{\mathscr{I}^{+}}_{\w \ell})  \;. \label{chi-scri-minus-2}  \eea \label{chi-scri-H-minus} \ees
The modes $f^K$ which are positive frequency on $H^{-}$ with respect to the Kruskal time coordinate can be expanded in terms of the $f^{H^{-}}$ modes.  The result
is given in~\eqref{alphaK-betaK}.

\section{Matching Coefficients}

\subsection{General Formulas}

In this section general formulas are derived for the matching coefficients used in an expansion of the modes of a massless minimally coupled scalar field for the {\it in} vacuum state in the collapsing null shell spacetime in terms of a complete set of modes in Schwarzschild spacetime in the region outside the past and future horizons.  These can be used in the computation of the stress-energy tensor, $\left\langle in \middle| T_{ab} \middle| in \right\rangle$ for the scalar field in the part of the collapsing null shell spacetime that is outside of the shell and outside of the event horizon.

The expansion of the {\it in} mode functions has the form
\bea
f^{in}_{\w \ell m} = \sum_{\ell'=0}^{\infty}\sum_{m'=-\ell'}^{\ell'} & \int_{0}^{\infty} d \w'
\Big[ A^{\mathscr{I^+}}_{\w \ell m\w' \ell' m'} f^{\mathscr{I}^+}_{\w'\ell' m'} + B^{\mathscr{I^+}}_{\w l m\w' \ell' m'} (f^{\mathscr{I}^+}_{\w' \ell' m'})^{*}
 \nonumber \\ & + A^{H^+}_{\w \ell m \w' \ell' m'} f^{H^+}_{\w' \ell' m'} + B^{H^+}_{\w \ell m\w'\ell' m'} (f^{H^+}_{\w' \ell' m'})^{*}\Big]\;. \label{General-in-modes}
\eea
The matching coefficients are found using the scalar product in~\eqref{scalar-products} and the orthonormality of the modes $f^{(\mathscr{I}^{+}, \, H^{+})}$ with respect to this scalar product.  The result is
\bes \bea
A^{(\mathscr{I^+}, H^+)}_{\w \ell m\w' \ell' m'} &=&( f^{\rm in}_{\w \ell m} , f^{(\mathscr{I}^+, H^+)}_{\w' \ell' m'}) \;, \\
B^{(\mathscr{I^+}, H^+)}_{\w \ell m\w' \ell' m'} &=& -( f^{\rm in}_{\w \ell m} , (f^{(\mathscr{I}^+, H^+)}_{\w' \ell' m'})^{*}) \;.
\eea \ees
For the Cauchy surface we consider, ~\eqref{scalar-products} reduces to integrals of the form
\bea
    \int du\int d\Omega r^2\overset{\leftrightarrow}{\partial_u}\;, \quad  \quad \int dv\int d\Omega r^2\overset{\leftrightarrow}{\partial_v}\;.
\eea
On the hypersurfaces where these integrals are computed, the following properties for spherical harmonics can be used
\bes \bea
    \int d\Omega Y_{lm}(\theta, \phi)Y_{l'm'}^{*}(\theta, \phi)&=&\delta_{l,l'} \delta_{m,m'},  \\
    \int d\Omega Y_{lm}(\theta, \phi)Y_{l'm'}(\theta, \phi)&=&(-1)^{m}\delta_{l,l'} \delta_{m,-m'} \;.
\eea \ees
As a result we can write
\bes \bea A^{(\mathscr{I^+}, H^+)}_{\w l m\w' l' m'} & =& \delta_{l,l'} \delta_{m,m'} A^{(\mathscr{I^+}, H^+)}_{\w \w' \ell}  \;, \label{A-short-def} \\
          B^{(\mathscr{I^+}, H^+)}_{\w l m\w' l' m'} & =& (-1)^m \delta_{l,l'} \delta_{m,-m'} B^{(\mathscr{I^+}, H^+)}_{\w \w' \ell}  \;, \label{B-short-def}
\eea \label{A-B-short-def} \ees
and
\be
f^{in}_{\w \ell m} =  \frac{Y_{\ell m}}{r \sqrt{4 \pi}} \int_{0}^{\infty} \frac{d \w'}{\sqrt{\w'}}
\Big[ A^{\mathscr{I^+}}_{\w \w' \ell} \psi^{\mathscr{I}^+}_{\w'\ell} + B^{\mathscr{I^+}}_{\w \w' \ell} (\psi^{\mathscr{I}^+}_{\w' \ell})^{*}
 + A^{H^+}_{\w \w' \ell} \psi^{H^+}_{\w' \ell} + B^{H^+}_{\w \w'\ell} (\psi^{H^+}_{\w' \ell})^{*}\Big]\;. \label{General-in-modes-2}
\ee
From this expression one can see that if, at small $\w'$, the matching coefficients go like $\frac{1}{\sqrt{\w'}}$ then there is an infrared divergence
in the integral and it is not obvious how to deal with it.  For this reason, we use integrations by parts in some of the computations of the matching coefficients below to avoid this difficulty.  For Schwarzschild spacetime in 4D, our results when substituted into~\eqref{General-in-modes-2}, do not give infrared divergences.

The contribution to the matching coefficients from the three segments of the Cauchy surface in Fig.~\ref{fig:Cauchy} are
\bes \bea
A^{(\mathscr{I^+}, H^+)}_{\w \w' l} & =& \left(A^{(\mathscr{I^+}, H^+)}_{\w \w' l} \right)_{H^{+}} + \left(A^{(\mathscr{I^+}, H^{+})}_{\w \w' l} \right)_{v_0}
   +  \left(A^{(\mathscr{I^+}, H^+)}_{\w \w' l} \right)_{\mathscr{I}^{-}} \;, \label{A-parts} \\
\left(A^{(\mathscr{I^+}, H^+)}_{\w \w' l} \right)_{H^{+}} &=&
-\frac{i}{4\pi\sqrt{\omega\omega'}}\int_{-\infty}^{v_{0}}dv\;\psi^{in}_{\omega l}(u = v_H,v) \overset{\leftrightarrow}{\partial_v}[\psi_{\omega' l}^{(\mathscr{I^+}, H^+)}(u_s = \infty,v)]^{*} \;, \\
\left(A^{(\mathscr{I^+}, H^{+})}_{\w \w' l} \right)_{v_0} &=& -\frac{i}{4\pi\sqrt{\omega\omega'}} \int_{-\infty}^{v_{H}}du\;\psi^{in}_{\omega l}(u,v_0) \overset{\leftrightarrow}{\partial_u}[\psi_{\omega' l}^{(\mathscr{I^+}, H^+)}(u_s(u),v_0)]^{*} \;, \label{A-I+-v0-1} \\
 \left(A^{(\mathscr{I^+}, H^+)}_{\w \w' l} \right)_{\mathscr{I}^{-}} &=& -\frac{i}{4\pi\sqrt{\omega\omega'}}\int_{v_{0}}^{\infty}dv\;\psi^{in}_{\omega l}(u =-\infty,v) \overset{\leftrightarrow}{\partial_v}[\psi_{\omega' l}^{(\mathscr{I^+}, H^+)}(u_s=-\infty,v) ]^{*}\;, \eea \ees
and
\bes \bea
B^{(\mathscr{I^+}, H^+)}_{\w \w' l} & =& \left(B^{(\mathscr{I^+}, H^+)}_{\w \w' l} \right)_{H^{+}} + \left(B^{(\mathscr{I^+}, H^{+})}_{\w \w' l} \right)_{v_0}
   +  \left(B^{(\mathscr{I^+}, H^+)}_{\w \w' l} \right)_{\mathscr{I}^{-}} \;, \label{B-parts} \\
\left(B^{(\mathscr{I^+}, H^+)}_{\w \w' l} \right)_{H^{+}} &=&
\frac{i}{4\pi\sqrt{\omega\omega'}}\int_{-\infty}^{v_{0}}dv\;\psi^{in}_{\omega l}(u = v_H,v) \overset{\leftrightarrow}{\partial_v} \psi_{\omega' l}^{(\mathscr{I^+}, H^+)}(u_s = \infty,v) \;, \\
\left(B^{(\mathscr{I^+}, H^{+})}_{\w \w' l} \right)_{v_0} &=& \frac{i}{4\pi\sqrt{\omega\omega'}} \int_{-\infty}^{v_{H}}du\;\psi^{in}_{\omega l}(u,v_0) \overset{\leftrightarrow}{\partial_u}\psi_{\omega' l}^{(\mathscr{I^+}, H^+)}(u_s(u),v_0) \label{B-I+-v0-1} \;, \\
 \left(B^{(\mathscr{I^+}, H^+)}_{\w \w' l} \right)_{\mathscr{I}^{-}} &=& \frac{i}{4\pi\sqrt{\omega\omega'}}\int_{v_{0}}^{\infty}dv\;\psi^{in}_{\omega l}(u=-\infty,v) \overset{\leftrightarrow}{\partial_v}\psi_{\omega' l}^{(\mathscr{I^+}, H^+)}(u_s=-\infty,v) \;. \eea \ees
Note that the modes $\psi_{\omega' l}^{(\mathscr{I^+})} $ vanish on $H^{+}$ so
\be \left(A^{\mathscr{I^+}}_{\w \w' l} \right)_{H^{+}} = \left(B^{\mathscr{I^+}}_{\w \w' l} \right)_{H^{+}} = 0 \;. \ee

As discussed in Sec.~\ref{sec:method-Tab}, it is necessary to specify $\psi^{\rm in}_{\w \ell}$ on the part of the future horizon with $v < v_0$.  This part is
causally disconnected from the region with $v > v_0$ outside the horizon so the only necessary constraint is that the mode functions should be continuous at the
point on the future horizon where $v = v_0$.  The simplest way to accomplish this is to take
\be \psi^{\rm in}_{\w \ell}(u = v_H, v) = \psi^{\rm in}_{\w \ell}(u = v_H, v_0) \;. \ee
With this choice it turns out to be useful to write the contribution to the matching coefficients from $H^+$ in the form
\bes \bea \left(A^{(\mathscr{I^+}, H^+)}_{\w \w' l} \right)_{H^{+}} &=&  \frac{i}{4 \pi \sqrt{\w \w'}} \psi^{\rm in}_{\w \ell}(v_H,v_0) e^{i \w' v_0}
  - \frac{i}{2 \pi} \sqrt{\frac{\w'}{\w}} \frac{e^{i \w' v_0}}{\w' - i \epsilon} \psi^{\rm in}_{\w \ell}(v_H,v_0)
   \;, \\
  \left(B^{(\mathscr{I^+}, H^+)}_{\w \w' l} \right)_{H^{+}} &=&  -\frac{i}{4 \pi \sqrt{\w \w'}} \psi^{\rm in}_{\w \ell}(v_H,v_0) e^{-i \w' v_0}
 + \frac{i}{2 \pi} \sqrt{\frac{\w'}{\w}} \frac{e^{i \w' v_0}}{\w' + i \epsilon} \psi^{\rm in}_{\w \ell}(v_H,v_0)  \;, \eea \ees
where for each integral an integration by parts has been done and an integrating factor $0 < \epsilon \ll 1$ has been included to make the integrals converge.

To avoid infrared divergences in~\eqref{General-in-modes-2} it is useful to subtract and then add back the quantity $e^{-i \w v_0}$ from $\psi^{\rm in}_{\w \ell}$
in~\eqref{A-I+-v0-1} and~\eqref{B-I+-v0-1}.  Then after integrations by parts the contributions from the surface $v = v_0$ can be written as
\bes \bea  \left(A^{ H^{+}}_{\w \w' l} \right)_{v_0} &=& - \frac{i}{4 \pi \sqrt{\w \w'} }  \psi^{\rm in}_{\w \ell}(u_H,v_0) e^{i \w' v_0} +  \frac{i}{4 \pi \sqrt{\w \w'}\, F_L^{*}(\w',\ell)}  e^{-i (\w-\w') v_0}  \nonumber \\
   & & + \frac{i}{2 \pi \sqrt{\w \w'} } \int_{-\infty}^{v_H} du \, \left[\partial_u \psi^{\rm in}_{\w \ell}(u,v_0) \right] \psi^{H^{+} *}_{\w' \ell}(u,v_0) \;, \\
   \left(B^{ H^{+}}_{\w \w' l} \right)_{v_0} &=&  \frac{i}{4 \pi \sqrt{\w \w'} }  \psi^{\rm in}_{\w \ell}(v_H,v_0) e^{-i \w' v_0} -  \frac{i}{4 \pi \sqrt{\w \w'}\, F_L(\w',\ell)}  e^{-i (\w+\w') v_0}  \nonumber \\
   & & - \frac{i}{2 \pi \sqrt{\w \w'} } \int_{-\infty}^{v_H} du \, \left[\partial_u \psi^{\rm in}_{\w \ell}(u,v_0) \right] \psi^{H^{+}}_{\w' \ell} (u,v_0)\;, \\
  \left(A^{ \mathscr{I}^{+}}_{\w \w' l} \right)_{v_0} &=& - \frac{i}{4 \pi \sqrt{\w \w'} } \frac{F^{*}_R(\w',\ell)}{F^{*}_L(\w', \ell)} e^{-i (\w - \w') v_0} \nonumber \\
 && + \frac{i}{2 \pi \sqrt{\w \w'}} \int_{-\infty}^{v_H} du \,  [\psi^{\rm in}_{\w \ell}(u, v_0) - e^{-i \w v_0}] \partial_u \psi^{\mathscr{I}^+ *}_{\w \ell} \;, \\
  \left(B^{ \mathscr{I}^{+}}_{\w \w' l} \right)_{v_0} &=&  \frac{i}{4 \pi \sqrt{\w \w'} } \frac{F_R(\w',\ell)}{F_L(\w', \ell)} e^{-i (\w + \w') v_0} \nonumber \\
 && - \frac{i}{2 \pi \sqrt{\w \w'}} \int_{-\infty}^{v_H} du \, \left[ \psi^{\rm in}_{\w \ell}(u, v_0) - e^{-i \w v_0} \right] \partial_u \psi^{\mathscr{I}^+}_{\w \ell} \;.
      \eea \ees

Since the spatial dependence of the modes is very simple on $\mathscr{I}^{-}$ it is possible to evaluate the integrals for the contributions to the matching parameters from there. After integrating by parts, we find
\bes \bea  \left(A^{ H^{+}}_{\w \w' l} \right)_{\mathscr{I}^{-}} &=& - \frac{i}{4 \pi \sqrt{\w \w'} \, F^{*}_L(\w',\ell)} e^{-i (\w-\w') v_0}
 + \frac{i}{2 \pi} \sqrt{\frac{\w'}{\w}} \frac{1}{F_L^{*}(\w',\ell)}\frac{e^{i(\w'-\w)v_0}}{\w'-\w+i \epsilon}  \;, \\
\left(B^{ H^{+}}_{\w \w' l} \right)_{\mathscr{I}^{-}} &=& \frac{i}{4 \pi \sqrt{\w \w'} \, F_L(\w',\ell)} e^{-i (\w+\w') v_0}
 - \frac{i}{2 \pi} \sqrt{\frac{\w'}{\w}} \frac{1}{F_L(\w',\ell)}\frac{e^{-i(\w+\w')v_0}}{\w'+\w-i \epsilon}  \;, \\
\left(A^{ \mathscr{I}^{+}}_{\w \w' l} \right)_{\mathscr{I}^{-}} &=& \frac{i}{4 \pi \sqrt{\w \w'}}\frac{F_R^{*}(\w',\ell)}{ F_L^{*}(\w',\ell)} e^{-i (\w-\w') v_0}
 - \frac{i}{2 \pi} \sqrt{\frac{\w'}{\w}} \frac{F_R^{*}(\w',\ell)}{F_L^{*}(\w',\ell)}\frac{e^{-i(\w-\w')v_0}}{\w'-\w+i \epsilon}  \;, \\
\left(B^{ \mathscr{I}^{+}}_{\w \w' l} \right)_{\mathscr{I}^{-}} &=& -\frac{i}{4 \pi \sqrt{\w \w'}}\frac{F_R(\w',\ell)}{ F_L(\w',\ell)} e^{-i (\w+\w') v_0}
 + \frac{i}{2 \pi} \sqrt{\frac{\w'}{\w}} \frac{F_R(\w',\ell)}{F_L(\w',\ell)}\frac{e^{-i(\w+\w')v_0}}{\w'+\w-i \epsilon}  \;.
\eea \ees

Combining these results together, the general formulas for the matching coefficients are
\bes \bea A^{ H^{+}}_{\w \w' l} &=& - \frac{i}{2 \pi} \sqrt{\frac{\w'}{\w}} \frac{e^{i \w' v_0}}{\w' - i \epsilon} \psi^{\rm in}_{\w \ell}(v_H,v_0)
   + \frac{i}{2 \pi} \sqrt{\frac{\w'}{\w}} \frac{1}{F_L^{*}(\w',\ell)}\frac{e^{i(\w'-\w)v_0}}{\w'-\w+i \epsilon}  \nonumber \\
   & & + \frac{i}{2 \pi \sqrt{\w \w'} } \int_{-\infty}^{v_H} du \, \left[\partial_u \psi^{\rm in}_{\w \ell}(u,v_0) \right] \psi^{H^{+} *}_{\w' \ell}(u,v_0) \;, \label{A-H-gen-mat}\\
      B^{ H^{+}}_{\w \w' l} &=& \frac{i}{2 \pi} \sqrt{\frac{\w'}{\w}} \frac{e^{-i \w' v_0}}{\w' + i \epsilon} \psi^{\rm in}_{\w \ell}(v_H,v_0)
      - \frac{i}{2 \pi} \sqrt{\frac{\w'}{\w}} \frac{1}{F_L(\w',\ell)}\frac{e^{-i(\w+\w')v_0}}{\w'+\w-i \epsilon} \nonumber \\
      & & - \frac{i}{2 \pi \sqrt{\w \w'} } \int_{-\infty}^{v_H} du \, \left[\partial_u \psi^{\rm in}_{\w \ell}(u,v_0) \right] \psi^{H^{+}}_{\w' \ell} (u,v_0)\;, \label{B-H-gen_mat}\\
     A^{ \mathscr{I}^{+}}_{\w \w' l} &=&  - \frac{i}{2 \pi} \sqrt{\frac{\w'}{\w}} \frac{F_R^{*}(\w',\ell)}{F_L^{*}(\w',\ell)}\frac{e^{-i(\w-\w')v_0}}{\w'-\w+i \epsilon} \nonumber \\  & &
       -\frac{i}{2 \pi \sqrt{\w \w'}} \int_{-\infty}^{v_H} du \, \left[  \psi^{\rm in}_{\w \ell}(u, v_0) - e^{-i \w v_0} \right] \partial_u \psi^{\mathscr{I}^+  *}_{\w' \ell} \;,  \label{A-I-gen-mat} \\
     B^{ \mathscr{I}^{+}}_{\w \w' l} &=&  \frac{i}{2 \pi} \sqrt{\frac{\w'}{\w}} \frac{F_R(\w',\ell)}{F_L(\w',\ell)}\frac{e^{-i(\w+\w')v_0}}{\w'+\w-i \epsilon} \nonumber \\
  & &     + \frac{i}{2 \pi \sqrt{\w \w'}} \int_{-\infty}^{v_H} du \, \left[ \psi^{\rm in}_{\w \ell}(u, v_0) - e^{-i \w v_0} \right] \partial_u \psi^{\mathscr{I}^+}_{\w' \ell} \;.
     \label{B-I-gen-mat} \eea \label{gen-mat}\ees

\subsection{Expansion of the Kruskal modes}

As discussed in Sec.~\ref{sec:method-Tab}, our method for renormalizing the stress-energy tensor involves subtracting the unrenormalized stress-energy tensor for the Unruh modes.  For this purpose the best way to do this is to express the modes $f^K_{\w \ell m}$ that are positive frequency on the past horizon with respect to the Kruskal time coordinate in terms of the modes $f^{H^{-}}_{\w \ell m}$ that are positive frequency with respect to the usual time coordinate $t_s$ on the past horizon. Then the relation~\eqref{chi-H-minus-2} can be used to express $f^K_{\w \ell m}$ in terms of the modes $f^{(\mathscr{I}^{+}\,H^{+})}_{\w \ell m}$.
The initial expansion can be written as
\be f^K_{\w \ell m} =  \int_0^\infty d \w' \, \left[  \alpha^K_{\w \w' \ell} f^{H^{-}}_{\w' \ell m} + \beta^K_{\w \w' \ell}   f^{H^{-}*}_{\w' \ell m} \right]  \;. \label{fK} \ee
 The Bogolubov coefficients can be obtained using the scalar product~\eqref{scalar-products}
   with a Cauchy surface consisting of the union of past null infinity and the past horizon in Schwarzschild spacetime.  Integrating over the angular coordinates one finds that
   the Bogolubov coefficients can be written in the form~\eqref{A-B-short-def} with $\alpha$ replacing $A$ and $\beta$ replacing $B$.
   Integrating the remaining integrals over $u_s$ by parts one finds that\footnote{This calculation was originally done in~\cite{BEC-2013} but note that there is a mistake in the results.  The expressions in that paper are missing a factor of $(4M)^{\pm i \w'}$. }
\bes \bea \alpha^K_{\w_K \w' \ell} &=& - \frac{i}{2 \pi} \sqrt{\frac{\w'}{\w_K}} (4 M)^{1+i 4 M \w'}  \int_{-\infty}^0 d U_K e^{-i \w_K U_K} (-U_K)^{-1 - i 4 M \w'} \nonumber \\
&=& \frac{1}{2 \pi} \sqrt{\frac{\w'}{\w_K}} \, (4M)^{1 +i 4 M \w'} \frac{\Gamma(\delta - i 4 M \w)}{ (-i \w_K + \epsilon)^{-i 4 M \w'}} \;, \label{alphaK} \\
\beta^K_{\w_K \w' \ell} &=&  \frac{i}{2 \pi} \sqrt{\frac{\w'}{\w_K}} (4 M)^{1-i 4 M \w'}  \int_{-\infty}^0 d U_K e^{-i \w_K U_K} (-U_K)^{-1 + i 4 M \w'} \nonumber \\
&=& \frac{1}{2 \pi} \sqrt{\frac{\w'}{\w_K}} \, (4M)^{1 -i 4 M \w'} \frac{\Gamma(\delta + i 4 M \w)}{ (-i \w_K + \epsilon)^{i 4 M \w'}} \;. \label{betaK}
\eea \label{alphaK-betaK} \ees
Here $\delta$ and $\epsilon$ are integrating factors with $0 < \delta \ll 1$ and $0 < \epsilon \ll 1$.  Note that the Bogolubov coefficients are independent of the value of $\ell$.  This is because the effective potential vanishes on $H^{-}$ which is the surface where the integrals are being computed.

\subsection{2D example}

In this section we will illustrate the matching for the case of a 2D spacetime which has a perfectly reflecting mirror at $r = 0$.  The metric
inside the shell is the flat space metric
\be
ds^2 = -dt^2 + dr^2    \;, \label{metric-flat-2D}
\ee
and the metric outside the shell is the Schwarzschild metric
\be
ds^2 = -\left(1-\frac{2M}{r} \right) dt_s^2
+ \left( 1-\frac{2M}{r} \right)^{-1} dr^2   \;. \label{metric-sch-2D}
\ee
The Penrose diagram is the same as in the 4D case as is the definition of the radial null coordinates $u$, $u_s$, and $v$ and the relation between $u$ and $u_s$.

The general form of the mode functions is
\be f_\w = \frac{\psi_\w}{\sqrt{4 \pi \w}}  \;.  \label{f-form-2D} \ee
There is no scattering for the massless minimally coupled scalar field modes in 2D so
\be E_R = F_L = 1\;, \qquad E_L = F_R = 0 \;. \label{scattering-coeff-2D} \ee

Inside the shell the {\it in} modes are
\be \psi^{\rm in}_\w =   e^{-i \w v} - e^{-i \w u} \;.  \label{psi-in-2D} \ee
In the region outside the shell the spacetime is the 2D version of Schwarzschild spacetime and the modes are
\bes \bea   \psi^{\mathscr{I}^{+}}_\w &=& \psi^{H^{-}}_\w = e^{- i \w u_s}  \;, \label{right-moving-2D} \\
           \psi^{H^{+}}_\w &=& \psi^{\mathscr{I}^{-}}_\w  = e^{-i \w v} \;. \label{left-moving-2D} \eea \label{sch-modes-2D} \ees

The expansion for the {\it in} modes is similar to the 4D case except there are no parameters $\ell$ and $m$ related to the spherical harmonics.  Thus
\bea
f^{in}_{\w} & =  & \int_{0}^{\infty} d \w'
\Big[ A^{H^+}_{\w \w' } f^{H^+}_{\w' } + B^{H^+}_{\w \w'} (f^{H^+}_{\w' })^{*}
   + A^{\mathscr{I^+}}_{\w \w'} f^{\mathscr{I}^+}_{\w'} + B^{\mathscr{I^+}}_{\w \w' } (f^{\mathscr{I}^+}_{\w' })^{*}\Big]\;.\label{2D-in-mode-expansion}
\eea
The matching coefficients are given by substituting~\eqref{scattering-coeff-2D},~\eqref{psi-in-2D}, and~\eqref{sch-modes-2D} into~\eqref{gen-mat}.  It is then easy to show that
\bea \left[ B^{H^{+}}_{\w,  \w'} f^{H^+\, *}_{\w' } \right]_{\w' \to - \w'} &=& - A^{H^{+}}_{\w,  \w'} f^{H^+}_{\w' } \;, \nonumber \\
     \left[ B^{\mathscr{I}^{+}}_{\w,  \w'} f^{\mathscr{I}^{+} \, *}_{\w' } \right]_{\w' \to - \w'} &=& - A^{ \mathscr{I}^{+}}_{\w,  \w'} f^{\mathscr{I}^{+}}_{\w' } \;,
   \label{B-w-minus-w} \eea
where the quantities on the right hand side are to be evaluated at $\w' < 0$.  As a result
\bea
f^{in}_{\w} & =  & \int_{-\infty}^{\infty} d \w'
\Big[   A^{H^+}_{\w \w' } f^{H^+}_{\w' } + A^{\mathscr{I^+}}_{\w \w'} f^{\mathscr{I}^+}_{\w'} \Big]\;. \label{2D-in-mode-expansion-2}
\eea
Because $\psi^{H^{+}}_{\w'}$ does not depend on $u$, the integral in~\eqref{A-H-gen-mat} is trivial to evaluate and one finds that
\be A^{H^+}_{\w \w'} = - \frac{i}{2 \pi} \sqrt{\frac{\w'}{\w}} \frac{e^{-i (\w - \w') v_0}}{\w' - i \epsilon}
   + \frac{i}{2 \pi} \sqrt{\frac{\w'}{\w}} \frac{e^{i(\w'-\w)v_0}}{\w'-\w+i \epsilon}
  \;. \label{A-H-2D} \ee

To see what the contribution to $f^{\rm in}_\w$ is from the $f^{H^{+}}_\w$ modes, first substitute ~\eqref{A-H-2D} into~\eqref{2D-in-mode-expansion} along with~\eqref{f-form-2D}
and~\eqref{psi-in-2D} with the result
\bea \left(f^{\rm in}_\w \right)_{H^{+}}  &=&  \frac{i e^{-i \w v_0}}{2 \pi \sqrt{4 \pi \w}} \int_{-\infty}^\infty d w' \; \left[ e^{i \w' (v_0 - v)} \left(-\frac{1}{\w' - i \epsilon}  +   \frac{1}{\w' - \w + i \epsilon} \right) \right] \nonumber \\
 & = &  \frac{ e^{-i \w v_0}}{\sqrt{4 \pi \w}}  \theta(v_0-v) + \frac{ e^{-i \w v}}{\sqrt{4 \pi \w}}  \theta(v-v_0) \;. \label{f-in-H+-2D}\eea

We next consider the contribution of the $f^{\mathscr{I}^{+}}$ modes.  The matching coefficient in~\eqref{A-I-gen-mat} is
\bea A^{\mathscr{I}^{+}}_{\w \w'} &=& -\frac{1}{2 \pi }\sqrt{\frac{\w'}{\w}} \int_{-\infty}^{v_H} du \, e^{-i \w u}  e^{i \w' u_s(u)} \frac{d u_s}{du} \nonumber \\
&=&
-\frac{1}{2 \pi} \sqrt{\frac{\w'}{\w}}  \int_{-\infty}^{v_H} du \, e^{-i (\w - \w') u}  \left(\frac{v_H-u}{4 M} \right)^{-i 4 M \w'} \left[ 1 + \frac{4M}{v_h-u} \right]  \;. \eea
Changing variables to $x = v_H -u$ and performing an integration by parts gives
\bea A^{\mathscr{I}^{+}}_{\w \w'} &=& \frac{i}{2 \pi} \sqrt{\w \w'} e^{-i(\w-\w') v_H} \frac{(4 M)^{1+i 4 M \w'}}{i (\w'-\w)+\epsilon} \int_0^\infty dx \; e^{i(\w-\w')x-\epsilon x}  \;
  x^{-i 4 M \w' -1 + \delta} \nonumber \\
  &=& \frac{i}{2 \pi} \sqrt{\w \w'} e^{-i(\w-\w') v_H} (4 M)^{1+i 4 M \w'} \frac{\Gamma(\delta - i 4 M \w')}{[i (\w'-\w)+ \epsilon]^{1-i4M\w'} }  \;. \label{Ascrp2D} \eea
Note that two integrating factors have been used with $0 < \epsilon \ll 1$ and $0 < \delta \ll 1$.

To find the contribution to $f^{\rm in}_\w$ from the $f^{\mathscr{I}^{+}}_\w$ modes, first substitute~\eqref{Ascrp2D}  into~\eqref{2D-in-mode-expansion} with the result
\bea \left(f^{\rm in}_\w\right)_{\mathscr{I}^{+}} &=& \frac{i 4 M \sqrt{\w}}{2 \pi \sqrt{4 \pi}} e^{-i \w v_H}
\int_{-\infty}^\infty d \w'   e^{i \w' (v_H-u_s)} (4 M)^{i 4 M \w'} \frac{\Gamma(\delta - i 4 M \w')}{[i (\w'-\w) + \epsilon]^{1-i 4 M \w'}} \;. \label{I1-a} \eea
Note that the denominator has an essential singularity in the upper half $\w'$ plane while the Gamma function has simple poles in the lower half plane at
 \be  \w' = - \frac{i \delta}{4 M} \;, \ee
  and
  \be \w' = - \frac{i n}{4 M} \;, \qquad n = 1, 2, \ldots \ee
In the complex plane at large $|\w'|$ Sterling's approximation gives
\be  \Gamma( -i 4 M \w') \approx \sqrt{2 \pi} e^{i 4 M \w'} e^{(-i 4 M \w' - 1/2) \log(-i 4 M \w')}   \;. \ee
Using the usual change of variables $\w' = R e^{i \theta}$, with $R > 0$, it is straight-forward to show that the dominant contribution to the integrand of $I_1$ in the large $R$ limit comes from the factor $e^{4 M R \sin \theta \, \log R}$ and therefore one must close in the lower half plane.  This means there is no contribution from the essential singularity but there is a contribution from each pole of the Gamma function.  At these poles it is straight-forward to show that
\be \Gamma(\delta -i 4 M \w) \to \frac{(-1)^n}{n! (n - i 4 M \w)}  \;, \qquad n = 0, 1, 2, \ldots  \ee
Then
\be  \left(f^{\rm in}_\w\right)_{\mathscr{I}^{+}}  = \frac{4Mi\sqrt{\w}}{\sqrt{4 \pi}} e^{- i \w v_H} \sum_{n=0}^\infty \frac{(-1)^n}{n!} (n-i 4 M \w)^{n-1}
    \left[\exp\left(\frac{(v_H-u_s)}{4 M} \right)\right]^n  \;. \label{finu} \ee
Because the general solutions to the 2D mode equation in Schwarzschild spacetime are of the form $\psi = g(u_s) + h(v)$ with $g$ and $h$ arbitrary functions, the exact solution for the {\it in} modes is
\be \left(f^{\rm in}_\w\right)_{\mathscr{I}^{+}} = - \frac{e^{-i \w u(u_s)}}{\sqrt{4\pi\w}}  = -\frac {e^{-i \w v_H}}{\sqrt{4\pi\w}} \, \exp \left\{ i 4 M \w\;  W\left[\exp\left(\frac{(v_H-u_s)}{4M}\right)\right] \right\} \;, \label{f(u)Lambert}\ee
where~\eqref{u-us} has been used and $W(z)$ is the Lambert W function.
  To make a comparison between ~\eqref{finu} and ~\eqref{f(u)Lambert}, one needs to write the latter in terms of a series. This has been done in~\cite{Corless}.
  An alternative derivation is given in Appendix~\ref{appendix-A}.  The result is
\be e^{- c W(z)}=\sum_{n=0}^\infty \frac {c (n+c)^{n-1}}{(n)!} (-z)^n\;. \label{final-Lambert}\ee
A detailed derivation of this expression is presented in Appendix A.
Taking $c=-4iM\w$ and $z=\exp\left(\frac{v_H-u_s}{4M}\right)$ in ~\eqref{final-Lambert}, one can see that~\eqref{f(u)Lambert} and ~\eqref{finu} are equivalent.

\subsection{Delta function potential}
\label{sec:delta-function}

In this section, we apply our matching method to the case where the potential term in ~\eqref{chi-eq-outside} is replaced by
\be V = \lambda \delta(r_*)  \;, \label{V-delta} \ee
with $\lambda$ a positive real constant.  This can serve as a model for the original potential which has a single peak and vanishes at the horizon and infinity.
The resulting mode equation can be solved analytically and the solutions are simple enough that the matching coefficients can be computed analytically. Some of these matching coefficients will be used to partially reconstruct the mode functions $f^{\rm in}_{\w \ell}$ in the case that $\ell = 0$.

For $\ell = 0$ in 4D the {\it in} modes inside the null shell take on the particularly simple form~\eqref{psi-in-0}.
In the region outside the shell the mode functions in the complete set with $\ell = 0$  have the general form
\be f^{(H^{+}, \mathscr{I}^{+})}_{\w' 0 0} = \frac{Y_{00}}{r \sqrt{4 \pi \w'}} \psi^{(H^{+}, \mathscr{I}^{+})}_{\w'0} \;, \qquad \psi^{(H^{+}, \mathscr{I}^{+})}_{\w'0} = e^{-i \w' t_s} \chi^{(H^{+}, \mathscr{I}^{+})}_{\w'0} \;.  \label{f-H-I-delta} \ee
The radial parts of the modes satisfy the following equation
\be
    \frac{d^{2}\chi}{dr_{*}^2}+(\omega^2-\lambda\delta(r_{*}))\chi=0 \;.
\label{DiracDeltaModeEquation} \ee
In the region where $r_*>0$, two linearly independent solutions are
\bes \bea
\chi_R &=& e^{i\omega r_{*}}\;, \label{RightMoving1}\\
\chi_L &=& e^{-i\omega r_{*}}\;. \label{LeftMoving1}
\eea \label{modes-delta-rs-gt-0} \ees
 For $r_*<0$, $\chi_R$ and $\chi_L$ can be expressed in the following way
\bes \bea
  \chi_{R}&=& E_R e^{i\omega r_{*}}+ F_R e^{-i\omega r_{*}}\;,  \label{RightMoving2}\\
  \chi_{L}&=& E_L e^{i\omega r_{*}}+ F_L e^{-i\omega r_{*}}\;,  \label{LeftMoving2}
\eea \label{modes-delta-rs-lt-0}\ees
where $E_R$ and $F_L$ are reflection and $F_R$ and $E_L$ are transmission coefficients.
Imposing the continuity of the mode function and discontinuity of its first derivative in the usual way at the spacelike curve $r_{*}=0$, the following analytic expressions are found for the scattering coefficients
\bes \bea
E_R &=& 1+\frac{i\lambda}{2\w}\; , \label{Dirac-ER}\\
F_R &=& -\frac{i\lambda}{2\w}\; ,\label{Dirac-FR}\\
E_L &=& F_R ^{*}=\frac{i\lambda}{2\w}\; ,\label{Dirac-EL}\\
F_L &=& E_R ^{*}=1-\frac{i\lambda}{2\w}\; .\label{Dirac-FL}
\eea \label{Scattering-Coefficients} \ees

Then the mode functions that we are using for the matching are
\bes \bea \psi^{H^{+}}_{\w'0}  &=& \theta(-r_*) \left[ e^{-i \w' v} + \frac{\frac{i \lambda}{2}}{\left(\w' - \frac{i \lambda}{2} \right)} e^{-i \w' u_s} \right]
  + \theta(r_*) \frac{\w'}{\left(\w' - \frac{i \lambda}{2} \right)} e^{-i \w' v} \;, \label{psi-H-delta} \\
  \psi^{\mathscr{I}^{+}}_{\w'0} &=& \theta(-r_*) \frac{\w'}{\left(\w' - \frac{i \lambda}{2} \right)} e^{-i \w' u_s}  + \theta(r_*) \left[  \frac{\frac{i \lambda}{2}}{\left(\w' - \frac{i \lambda}{2} \right)} e^{-i \w' v} + e^{-i \w' u_s} \right] \;. \label{psi-I-delta} \eea \label{psi-delta} \ees

To verify that the matching coefficients can be used to reconstruct the original mode functions for the case $\ell = 0$ it is useful to break them up into
contributions that come from the term proportional to  $e^{-i \w v}$ in~\eqref{psi-in-0} and the term proportional to $-e^{-i \w u}$.  In what follows we compute
the matching coefficients for both terms but then focus only those that come from the term proportional to $e^{-i \w v}$.
Substituting~\eqref{psi-delta}, and~\eqref{Scattering-Coefficients} into~\eqref{gen-mat} one finds the matching coefficients
\bes \bea
A^{H^{+}}_{\w \w' 0} &=& (A^{H^{+}}_{\w \w' 0})_v  + (A^{H^{+}}_{\w \w' 0})_u  \;, \nonumber \\
(A^{H^{+}}_{\w \w' 0})_v &=& - \frac{i}{2 \pi} \sqrt{\frac{\w'}{\w}} \frac{e^{i \w' v_0}}{\w' - i \epsilon} e^{-i \w v_0}
   + \frac{i}{2 \pi} \sqrt{\frac{\w'}{\w}} \frac{1}{F_L^{*}(\w',\ell)}\frac{e^{i(\w'-\w)v_0}}{\w'-\w+i \epsilon} \;, \label{AHv-delta} \\
 (A^{H^{+}}_{\w \w' 0})_u &=&   \frac{i}{2 \pi} \sqrt{\frac{\w'}{\w}} \frac{e^{i \w' v_0}}{\w' - i \epsilon} e^{-i \w v_H}
    - \frac{1}{2 \pi} \sqrt{\frac{\w}{ \w'} } \int_{-\infty}^{v_H} du \,  e^{-i w u}  \left[ \theta(r_{*}) \frac{\w'}{\w' + \frac{i \lambda}{2}} e^{i \w' v_0} \right.  \nonumber \\
 & &  \left. \qquad    + \, \theta(-r_{*}) \left( e^{i \w' v_0}
     - \frac{\frac{i \lambda}{2}}{\w' + \frac{i \lambda}{2}} e^{i \w' u_s(u)} \right) \right] \;, \label{AHu-delta}
\eea \ees
\bes \bea
A^{\mathscr{I}^{+}}_{\w \w' 0} &=& (A^{\mathscr{I}^{+}}_{\w \w' 0})_v  + (A^{\mathscr{I}^{+}}_{\w \w' 0})_u  \;, \nonumber \\
(A^{\mathscr{I}^{+}}_{\w \w' 0})_v &=&  - \frac{i}{2 \pi} \sqrt{\frac{\w'}{\w}} \frac{\frac{i \w' \lambda}{2}}{\w' + \frac{i \lambda}{2}}\frac{e^{-i(\w-\w')v_0}}{\w'-\w+i \epsilon}\;,  \label{AIv-delta}   \\
  (A^{\mathscr{I}^{+}}_{\w \w' 0})_u &=&  -\frac{1}{2 \pi} \sqrt{\frac{\w'}{ \w}} \int_{-\infty}^{v_H} du  \frac{du_s(u)}{du} \, e^{-i \w u} e^{i \w' u_s(u)} \left[ \theta(r_{*})  +
  \theta(-r_{*}) \frac{\w'}{\w'-\frac{i \lambda}{2}}  \right]
   \;.
  \label{AIu-delta} \eea \ees
Note that the relations~\eqref{B-w-minus-w}  are satisfied by these matching coefficients so the relation~\eqref{2D-in-mode-expansion-2} also holds. Thus
 \bea
 (f^{in}_{\w 0 0})_v &=&  \int_{-\infty}^\infty d \w'   \left[ (A^{H^{+}}_{\w \w' 0})_v  f^{H^+}_{\w'00} +
    (A^{\mathscr{I}^{+}}_{\w \w' 0})_v  f^{\mathscr{I}^+}_{\w'00} \right] \;.
    \label{in-modes-v-part}
    \eea
Substituting~\eqref{AHv-delta},~\eqref{AIv-delta}, and~\eqref{psi-delta} into~\eqref{in-modes-v-part} gives after some algebra
\bes \bea  (f^{in}_{\w00})_v &=& \frac{Y_{00}}{r \sqrt{4 \pi \w}} \left[\theta(-r_*) I_1 + \theta(r_*) I_2 \right] \;, \nonumber \\
   I_1 &=& - \frac{i}{2 \pi } e^{-i \w v_0} \int_{-\infty}^\infty d \w' \left[ \frac{\frac{i \lambda}{2} e^{i \w'(v_0-u_s)}}{\left(\w' - \frac{i \lambda}{2} \right)}
     + \frac{e^{i \w'(v_0-v)}}{(\w'-i \epsilon)} - \frac{\w' e^{i \w' (v_0-v)}}{\left(\w' + \frac{i \lambda}{2} \right) \left( \w' - \w + i \epsilon \right)} \right] \nonumber \\
     &=&  \theta(v_0-u_s) e^{-i \w v_0} \left[e^{-\frac{\lambda}{2}(v_0-u_s)} - 1 \right] + \theta(v_0-v) e^{-i \w v_0} \nonumber \\
   & &     + \frac{ \theta(v-v_0)}{\left( \w + \frac{i \lambda}{2} \right)} \left[ \frac{i \lambda}{2} e^{-i \w v_0} e^{- \frac{\lambda}{2}(v-v_0)} + \w e^{- i \w v} \right]\;, \\
   I_2 &=& - \frac{i}{2 \pi } e^{-i \w v_0} \int_{-\infty}^\infty d \w' \left[ \frac{\w' e^{i \w'(v_0-v)}}{(\w'-i \epsilon)\left(\w' - \frac{i \lambda}{2} \right)}
   -  \frac{ e^{i \w'(v_0-v)}}{\w'-\w + i \epsilon} + \frac{\frac{i \lambda}{2} e^{i \w'(v_0-u_s)}}{(\w'-i\epsilon)\left(\w' - \frac{i \lambda}{2} \right)} \right] \nonumber \\
   &=&   \theta(v_0-v) e^{-i \w v_0} e^{-\frac{\lambda}{2}(v_0-v)} + \theta(v-v_0) e^{-i \w v}  \nonumber \\
    & & - \theta(u_s-v_0) \frac{ \frac{i \lambda}{2}}{\left( \w + \frac{i \lambda}{2} \right)} \left[
   - e^{-i \w v_0} e^{-\frac{\lambda}{2}(u_s-v_0)} + e^{-i \w u_s} \right]\;.
 \eea \label{f-in-v}\ees
It is easy to verify that~\eqref{f-in-v} gives the correct values for $(f^{in}_{\w00})_v$ on the future horizon for $v \le v_0$, on the null shell surface $v = v_0$, and  on past null infinity for $v \ge v_0$.

\subsection{Partial analytic results for the matching coefficients in 4D for $\ell = 0$}

Because of the simple form of the {\it in} modes for $\ell = 0$ inside the null shell~\eqref{psi-in-0}, it is possible to compute the matching coefficients
for the $e^{-i \w v}$ part analytically.  To do so we begin by
substituting~\eqref{psi-in-0} into~\eqref{gen-mat} with the result
\bes \bea A^{ H^{+}}_{\w \w' 0} &=& - \frac{i}{2 \pi} \sqrt{\frac{\w'}{\w}} \frac{e^{i \w' v_0}}{\w' - i \epsilon} (e^{-i \w v_0} - e^{-i \w v_H})
   + \frac{i}{2 \pi} \sqrt{\frac{\w'}{\w}} \frac{1}{F_L^{*}(\w',0)}\frac{e^{i(\w'-\w)v_0}}{\w'-\w+i \epsilon}  \nonumber \\
   & & - \frac{1}{2 \pi} \sqrt{\frac{\w }{\w'}}  \int_{-\infty}^{v_H} du \,  \w e^{-i \w u}  \psi^{H^{+} *}_{\w' \ell}(u,v_0) \;, \label{A-H-ell-0-mat}\\
      B^{ H^{+}}_{\w \w' 0} &=& \frac{i}{2 \pi} \sqrt{\frac{\w'}{\w}} \frac{e^{-i \w' v_0}}{\w' + i \epsilon} (e^{-i \w v_0} - e^{-i \w v_H})
      - \frac{i}{2 \pi} \sqrt{\frac{\w'}{\w}} \frac{1}{F_L(\w',0)}\frac{e^{-i(\w+\w')v_0}}{\w'+\w-i \epsilon} \nonumber \\
      & & + \frac{1}{2 \pi} \sqrt{\frac{\w}{ \w'}}  \int_{-\infty}^{v_H} du \,e^{-i \w u} \psi^{H^{+}}_{\w' \ell} (u,v_0)\;, \label{B-H-ell-0-mat}\\
     A^{ \mathscr{I}^{+}}_{\w \w' 0} &=&  - \frac{i}{2 \pi} \sqrt{\frac{\w'}{\w}} \frac{F_R^{*}(\w',0)}{F_L^{*}(\w',0)}\frac{e^{-i(\w-\w')v_0}}{\w'-\w+i \epsilon}
       +\frac{i}{2 \pi} \sqrt{\frac{\w }{\w'}} \int_{-\infty}^{v_H} du \, e^{-i \w u} \partial_u \psi^{\mathscr{I}^+  *}_{\w \ell} \;,  \label{A-I-ell-0-mat} \\
     B^{ \mathscr{I}^{+}}_{\w \w' 0} &=&  \frac{i}{2 \pi} \sqrt{\frac{\w'}{\w}} \frac{F_R(\w',0)}{F_L(\w',0)}\frac{e^{-i(\w+\w')v_0}}{\w'+\w-i \epsilon}
     - \frac{i}{2 \pi} \sqrt{\frac{\w}{ \w'}} \int_{-\infty}^{v_H} du \,e^{-i \w u} \partial_u \psi^{\mathscr{I}^+}_{\w \ell} \;.
     \label{B-I-ell-0-mat} \eea \label{ell-0-mat}\ees

Note that the integrals have to be computed numerically because the mode functions in Schwarzschild spacetime must be computed numerically.
However, because of the simple form that $\psi^{\rm in}_{\w 0}$ takes it is possible to separate the matching coefficients into separate matching
coefficients for the part that goes like $e^{- i \w v}$ inside the null shell and the part that goes like $e^{-i \w u}$ there.  The matching coefficients
for $e^{-i \w v}$ don't depend on the integrals.  In what follows we focus on these matching coefficients.  Examination of~\eqref{ell-0-mat} gives for these coefficients
\bes \bea \left(A^{ H^{+}}_{\w \w' 0}\right)_v &=& - \frac{i}{2 \pi} \sqrt{\frac{\w'}{\w}} \frac{e^{i (\w'-\w) v_0}}{\w' - i \epsilon}
   + \frac{i}{2 \pi} \sqrt{\frac{\w'}{\w}} \frac{1}{F_L^{*}(\w',0)}\frac{e^{i(\w'-\w)v_0}}{\w'-\w+i \epsilon}  \nonumber \\
   & & \;, \label{A-H-ell-0-v-mat}\\
      \left(B^{ H^{+}}_{\w \w' l}\right)_v &=& \frac{i}{2 \pi} \sqrt{\frac{\w'}{\w}} \frac{e^{-i (\w'+\w) v_0}}{\w' + i \epsilon}
   - \frac{i}{2 \pi} \sqrt{\frac{\w'}{\w}} \frac{1}{F_L(\w',0)}\frac{e^{-i(\w+\w')v_0}}{\w'+\w-i \epsilon} \nonumber \\
      & & \;, \label{B-H-ell-0-v-mat}\\
     \left(A^{ \mathscr{I}^{+}}_{\w \w' l}\right)_v &=&  - \frac{i}{2 \pi} \sqrt{\frac{\w'}{\w}} \frac{F_R^{*}(\w',0)}{F_L^{*}(\w',0)}\frac{e^{-i(\w-\w')v_0}}{\w'-\w+i \epsilon}
      \;,  \label{A-I-ell-0-v-mat} \\
     \left(B^{ \mathscr{I}^{+}}_{\w \w' l}\right)_v &=&  \frac{i}{2 \pi} \sqrt{\frac{\w'}{\w}} \frac{F_R(\w',0)}{F_L(\w',0)}\frac{e^{-i(\w+\w')v_0}}{\w'+\w-i \epsilon}
     \;.
     \label{B-I-ell-0-v-mat} \eea \label{ell-0-v-mat}\ees

These matching coefficients can be used to reconstruct the part of the mode function which goes like $e^{-i \w v}$ inside the shell by substituting the expressions into~\eqref{General-in-modes}.  To check them we shall compute the resulting integral on $H^{+}$.  Recall that we are working in
the exact Schwarzschild spacetime rather than the null shell spacetime when we do the matching.  The same applies to the reconstruction.  Thus the results for the reconstruction for which $v \ge v_0$ also apply to the null shell spacetime,  but the results for $v < v_0$ do not apply to the null shell spacetime.

Recall that the modes $f^{\mathscr{I}^{+}}$ vanish on $H^{+}$.
\bes \bea \left(f^{\rm in}_{\w 0 0} \right)_v &=& \frac{Y_0^0}{r \sqrt{4 \pi \w}} (I_1 + I_2)\;,  \nonumber \\
   I_1 &=&  - \frac{i}{2 \pi} e^{-i \w v_0} \int_0^\infty d \w' \,\left[ \frac{e^{i(v_0-v)\w'}}{\w'-i \epsilon} -  \frac{e^{-i(v_0-v)\w'}}{\w'+i \epsilon} \right]\;, \label{I1} \\
   I_2 &=& \frac{i}{2 \pi} e^{-i \w v_0} \int_0^\infty d \w' \, \left[\frac{1}{F_L^{*}(\w',0)} \frac{e^{-i \w'(v-v_0)}}{\w'-\w + i \epsilon}
   - \frac{1}{F_L(\w',0)} \frac{e^{i \w'(v-v_0)}}{\w'+\w - i \epsilon} \right] \;. \label{I2}
 \eea \label{fin-ell-0-v}\ees
If in the second term of $I_1$ a change of variables is made so that $\w' \to - \w'$ then one finds that
\be I_1 =   - \frac{i}{2 \pi} e^{-i \w v_0} \int_{-\infty}^\infty d \w' \frac{e^{i(v_0-v)\w'}}{\w'-i \epsilon} = e^{-i \w v_0}\, \theta(v_0-v) \;,\ee
with $\theta$ the step function.  It is thus clear that the initial data on $H^{+}$ for $-\infty < v < v_0$ does not affect the mode functions on the
part of the future horizon for which $v_0 < v < \infty$.

It can be shown from the properties of the scattering coefficients given in~\cite{rigorous}, that $F_L(\w')=F^{*}_L(-\w')$. Using this identity and changing the variable of integration in the second integral in the same way as was done for $I_1$, one obtains
\be
(f^{in}_{\w 0 0})_v =\frac{e^{-i\w v_0}}{\sqrt{4\pi\w}}\theta(v_{0}-v)+\frac{i}{2\pi \sqrt{4\pi\w}}\int_{-\infty}^{\infty} d\w' \frac{e^{-i\w'v}}{
F^*_{L}(\w')}\frac{e^{iv_{0}(\w'-\w)}}{\w'-\w+i\epsilon} \;.\ee
To compute this integral using complex integration techniques one must know the singularity structure of $\frac{1}{F_L^{*}}$ which is difficult since this scattering coefficient
must be computed numerically.  However, one can at least test whether it has one or more singularities in the complex plane by assuming it does not and computing the integral.
We'll call the result $f^{\rm test}$ because there is no guarantee that this method will give the correct answer.
The result of such an integration is
\[ f^{\rm test} =\frac{e^{-i\w v_0}}{\sqrt{4\pi\w}}\theta(v_{0}-v)+\frac{e^{-i\w v}}{ F_{L}^{*}(\w)\sqrt{4\pi\w}}\theta(v-v_{0})\;.
\]
Here complex integration has been performed using a contour in the lower half of the complex plane.
It is obvious that at $v=v_{0}$, the continuity condition for $(f^{in}_{\w 0 0})_v$ is not satisfied so $f^{\rm test} \ne (f^{in}_{\w 0 0})_v$
which implies that $\frac{1}{F_{L}^{*}(\w')}$ has one or more singularities in the complex plane.

\begin{figure}[h]
\centering
\includegraphics [totalheight=0.27\textheight]{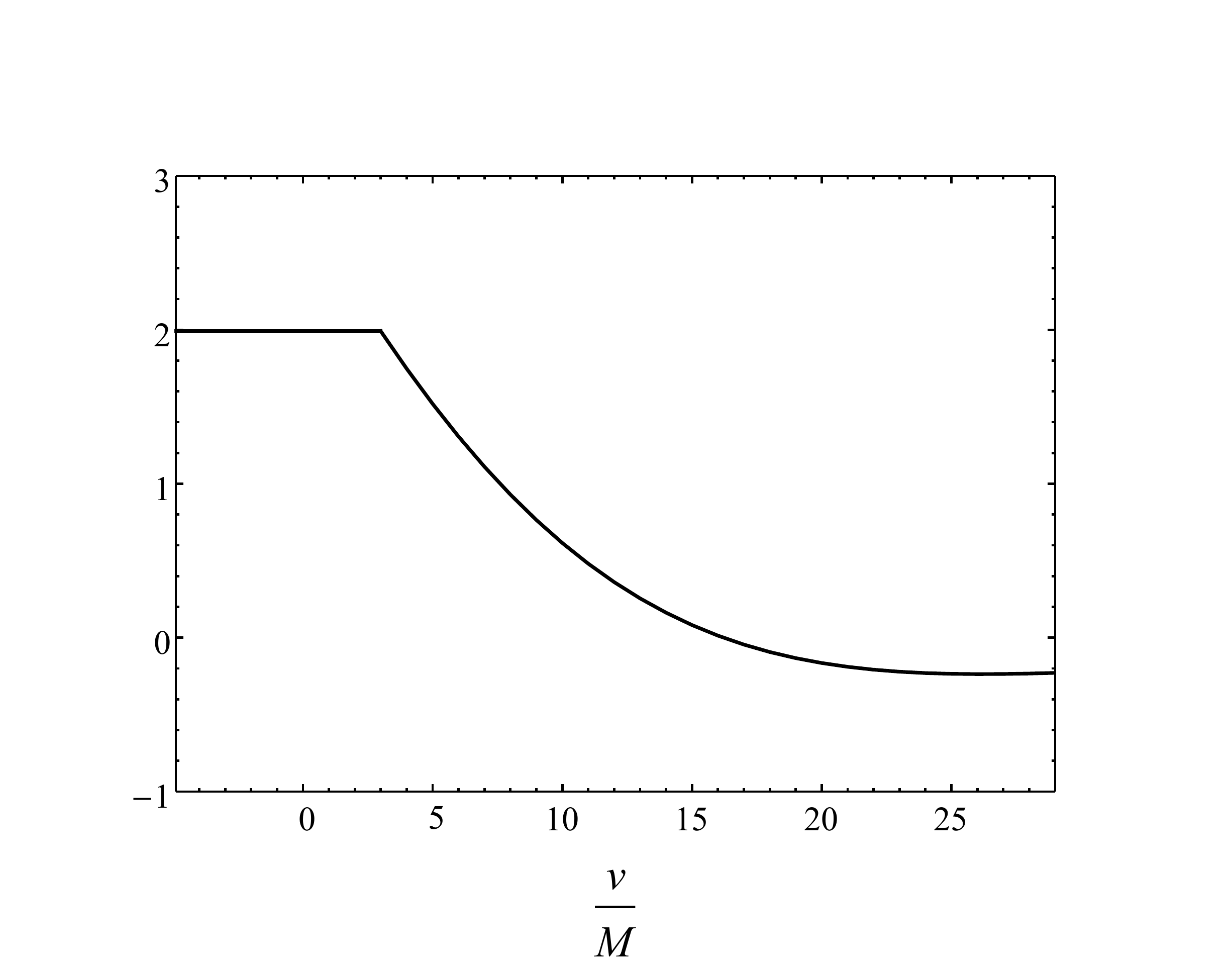}
\includegraphics [totalheight=0.27\textheight]{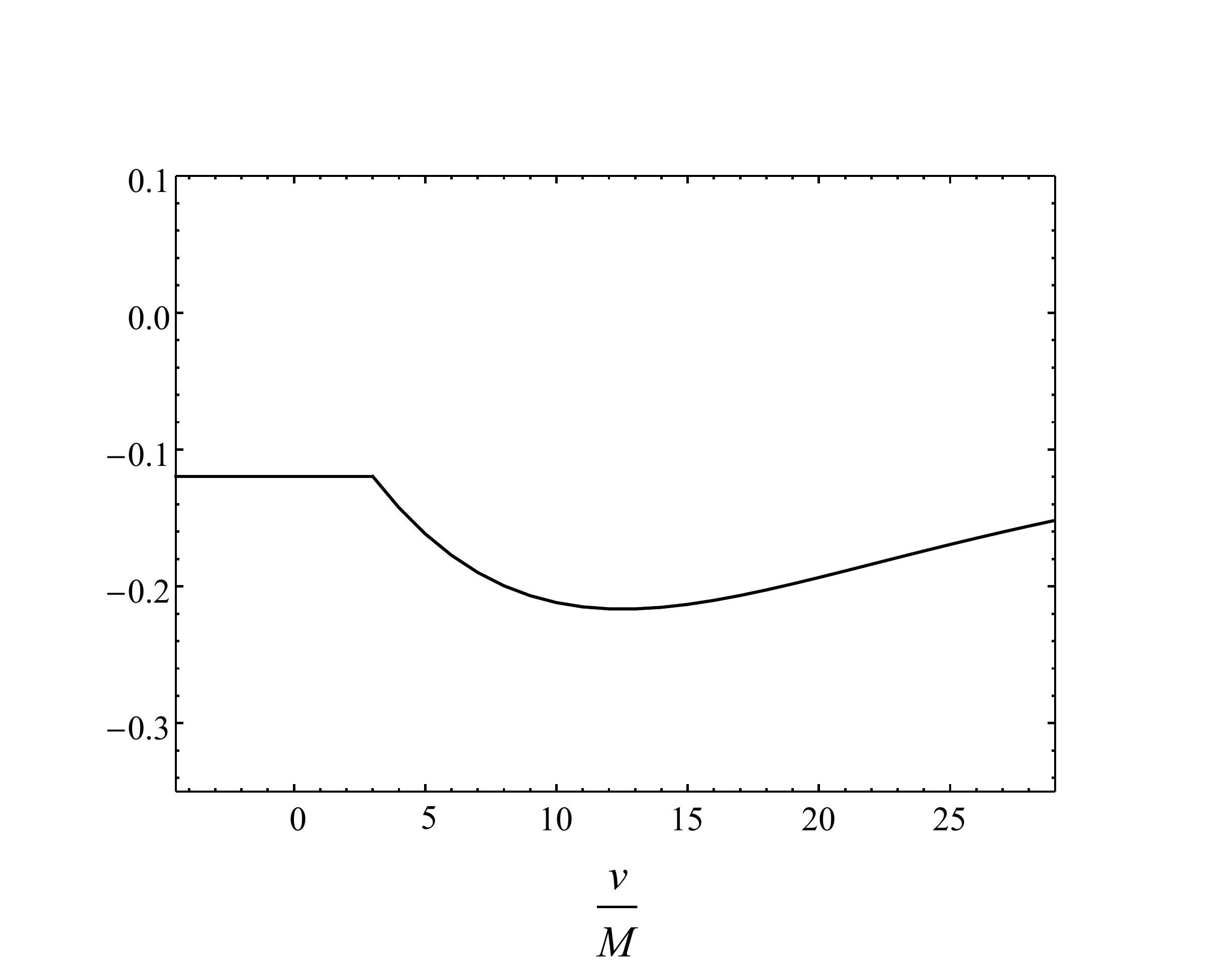}
\caption{The real (left) and the imaginary (right) parts of  $\sqrt{\frac{4 \pi}{M}} (f^{\rm in}_{\w 00})_v $ on the future horizon have been plotted. In both plots, $M\omega = 0.02$ and $\frac{v_0}{M}=3$. The plots clearly show that $(f^{\rm in}_{\w 00})_v$ is continuous at $v = v_0$.}
\label{fig:I2}
\end{figure}

Alternatively one work with $I_2$ in the form~\eqref{I2}, use the relation $(\w'\mp\w \pm i\epsilon)^{-1} = \mp i \pi \delta(\w'\mp\w) + (\w' \mp w)^{-1}$, and compute the principle value parts of the integral numerically for $v > v_0$.    This has been done and the result is shown in Fig.~\ref{fig:I2}.  It is clear from the plots in this figure that on the future horizon $(f^{\rm in}_{\w 0 0})_v$ is continuous at $v = v_0$.

\section{Stress-Energy Tensor}

\subsection{Method in 4D}
\label{sec-4D-method}

For the massless minimally coupled scalar field the classical stress-energy tensor in a general curved spacetime is given in~\eqref{Tab-class} and a renormalization expression for $\la T_{ab} \ra$ is given in~\eqref{Tab-Unruh-sub}.
To compute $\la T_{ab} \ra$ using~\eqref{Tab-Unruh-sub} it is useful to begin with the points split and to write the stress-energy tensor in terms of derivatives of the
Hadamard Green's function
\be  G^{(1)}(x,x') = \la \{ \phi(x), \phi(x') \}\ra  \;, \label{G1-def} \ee
Then adopting the notation
\be \Delta G^{(1)}(x,x') = \la {\rm in}| \{ \phi(x), \phi(x') \} | {\rm in}\ra - \la {U}| \{ \phi(x), \phi(x') \} | {U}\ra \;, \label{Delta-G-def} \ee
with $|{\rm in} \ra$ representing the {\it in} vacuum state and $|U \ra$ the Unruh state.  The corresponding difference in the stress-energy tensors is then
\be  \Delta \la T_{ab} \ra = \frac{1}{4} \lim_{x' \to x} \left[ \left( g_a^{c'} \Delta G^{(1)}_{;c'; b} + g_b^{c'}  \Delta G^{(1)}_{;a; c'} \right)  - \, g_{a b}\, g^{c d'} \Delta G^{(1)}_{;c ;d'} \right]  \;. \label{Delta-Tab-1} \ee
Here the quantity $g_a^{b'}$ parallel transports a vector from $x'$ to $x$ and is called the bivector of parallel transport~\cite{christensen-76}.
To leading order when the point separation is small
\be g_a^{b'} = g_a^b = \delta_a^b  \;. \label{gabp-leading}\ee
The subleading orders all vanish in the limit $x' \to x$.
Since there are no ultraviolet divergences in the quantity $\Delta \la T_{ab} \ra$ one can use~\eqref{gabp-leading} in~\eqref{Delta-Tab-1} with the result
\be  \Delta \la T_{ab} \ra =   \frac{1}{4}  \left[\lim_{x' \to x} \left( \Delta G^{(1)}_{;a' ;b} +   \Delta G^{(1)}_{;a ;b'} \right)  - g_{a b}\, g^{c d}  \lim_{x' \to x} \Delta G^{(1)}_{;c ;d'} \right]  \;, \label{Delta-Tab-2} \ee
where a slight abuse of notation has been used for the implied sum over $d$ and $d'$ in the last term.  It is important to note that this expression is valid in both two and four dimensions.

Expanding the field in terms of modes as in~\eqref{Phi-expansion} one finds for the {\it in} modes
that
\be   \la 0 \, {\rm in} |  \{ \phi(x), \phi(x') \} |0 \, {\rm in} \ra = \sum_{\ell = 0}^\infty \sum_{m = - \ell}^\ell \int_0^\infty d \w \; [f^{\rm in}_{\w \ell m}(x) (f^{\rm in}_{\w \ell m}(x'))^{*} + f^{\rm in}_{\w \ell m}(x') (f^{{\rm in}}_{\w \ell m}(x))^{*}]  \;.  \label{G1-in} \ee
The Unruh state in Schwarzschild spacetime consists of modes that are positive frequency with respect to the usual time coordinate on $\mathscr{I}^{-}$ along with modes that are positive frequency with respect to the Kruskal time coordinate on $H^{-}$ so that
\bea   \la U |  \{ \phi(x), \phi(x') \} |U \ra &=& \sum_{\ell = 0}^\infty \sum_{m = - \ell}^\ell \left\{ \int_0^\infty d \omega_K \, [f^{K}_{\w_K \ell m}(x) (f^{K}_{\w_K \ell m}(x'))^{*} + f^{K}_{\w_K \ell m}(x') (f^{{K}}_{\w_K \ell m}(x))^{*}] \right.  \nonumber \\
  && \; \left.  + \,  \int_0^\infty d \w \; [f^{\mathscr{I}^{-}}_{\w \ell m}(x) (f^{\mathscr{I}^{-}}_{\w \ell m}(x'))^{*} + f^{\mathscr{I}^{-}}_{\w \ell m}(x') (f^{{\mathscr{I}^{-}}}_{\w \ell m}(x))^{*}]  \right\}  \;.  \label{G1-U}  \eea

The next step is to find expansions for these two-point functions in terms of the complete set of modes $f^{(\mathscr{I}^{+}, H^{+})}$ that we are using.
For $ \la 0 \, {\rm in} |  \{ \phi(x), \phi(x') \} |0 \, {\rm in} \ra$ one can substitute~\eqref{General-in-modes} into~\eqref{G1-in}.  This results in integrals of the form
\bea  &&\sum_{\ell = 0}^\infty \sum_{m = - \ell}^\ell \int_0^\infty d \w \; \int_0^\infty d \w_1  \; \int_0^\infty d\w_2 \left\{ \left[  A^{(H^{+}, \mathscr{I}^{+})} f^{(H^{+}, \mathscr{I}^{+})} +  B^{(H^{+}, \mathscr{I}^{+})} (f^{(H^{+}, \mathscr{I}^{+})})^{*} \right] \right. \nonumber \\
&& \qquad \qquad \left.  \times \left[  (A^{(H^{+}, \mathscr{I}^{+})})^{*} (f^{(H^{+}, \mathscr{I}^{+})})^{*} +  (B^{(H^{+}, \mathscr{I}^{+})})^{*} f^{(H^{+}, \mathscr{I}^{+})} \right] \right\}
 \;, \label{triple-integral-form} \eea
where the subscripts on the matching coefficients and mode functions have been suppressed.
For $\la U |  \{ \phi(x), \phi(x') \} |U \ra$ one can first substitute~\eqref{fK} and~\eqref{alphaK-betaK} into~\eqref{G1-U} to obtain an expression in terms of
$f^{(H^{-},\mathscr{I}^{-})}$.  The relationship between these modes and $f^{(\mathscr{I}^{+}, H^{+})}$ can be deduced from~\eqref{chi-H-scri-def} and used to obtain an expression for $\la U |  \{ \phi(x), \phi(x') \} |U \ra$ that depends only on $f^{(\mathscr{I}^{+}, H^{+})}$.

\subsection{2D Example}
\label{sec:Tab-2D}

In this section the method discussed above is tested by using it to computing the stress-energy tensor for the scalar field in the corresponding 2D spacetime where the answer is known.  The computation will be done in the region $v > v_0$ outside the null shell and outside the horizon.
From~\eqref{f-in-H+-2D} it is clear that for $v > v_0$ the contribution from the $f^{H^{+}}_{\w'}$ modes to $f^{\rm in}_\w$ is
\be \left(f^{\rm in}_\w \right)_{H^{+}} = \frac{e^{-i \w v}}{\sqrt{4 \pi \w}}  = f^{H^{+}}_{\w}  \;. \label{f-in-H+-2} \ee
Thus
\bea f^{\rm in}_\w = f^{H^+}_\w +  \int_0^\infty d\w \; [A^{\mathscr{I}^{+}}_{\w \w'} f^{\mathscr{I}^{+}}_{\w'}  + B^{\mathscr{I}^{+}}_{\w \w'} (f^{\mathscr{I}^{+}}_{\w'})^{*}]\;, \label{fin-A-B}
\eea
with $A^{\mathscr{I}^{+}}_{\w \w'}$ given in~\eqref{Ascrp2D}. Using the relation $\Gamma(x) = \frac{\Gamma(1+x)}{x}$ one obtains the form used for the numerical computations
\bea
A^{\mathscr{I}^{+}}_{\w \w'} &=& -\frac{1}{2 \pi} \sqrt{\frac{\w}{ \w'}} (4M)^{i 4M \w'} e^{-i(\w-\w') v_H} \frac{\Gamma(1-i4M\w')}{[-i(\w-\w')+\epsilon]^{1-i4M\w'}}\;.  \label{AI2}
\eea
Then, using the relations~\eqref{B-w-minus-w} one finds
\be B^{\mathscr{I}^{+}}_{\w \w'} =  \frac{1}{2 \pi} \sqrt{\frac{\w}{ \w'}} (4M)^{-i 4M \w'} e^{-i(\w+\w') v_H} \frac{\Gamma(1+i4M\w')}{[-i(\w+\w')+\epsilon]^{1+i4M\w'}}  \;. \label{BI2}  \ee
In what follows the superscript $\mathscr{I}^+$ on the matching coefficients A and B will be suppressed.

Next, with the aim of finding the components of the stress-energy tensor using~\eqref{Delta-Tab-2}, we construct the Hadamard form of Green's function which in 2D is
\bea
 G^{(1)}(x,x') = \int_0^\infty d \w \; [f^{in}_\w(x) f^{in \;*}_{w}(x') + f^{in \;*}_\w(x) f^{in}_{w}(x') ]\;.
  \label{Hadamard}
 \eea
Substituting ~\eqref{fin-A-B} into ~\eqref{Hadamard} gives
 \bea
 G^{(1)}(x,x') &=& \int_0^\infty d \w \; \Big\{ \Big[f^{H^{+}}_\w(x) +  \int_0^\infty d\w_1 \; [A_{\w \w_1} f^{\mathscr{I}^{+}}_{\w_1}(x)  + B_{\w \w_1} f^{\mathscr{I}^{+}\;*}_{\w_1}(x)]\Big].  \nonumber \\
 & & \left.   \Big[f^{H^{+} \;*}_\w(x') +  \int_0^\infty d\w_2 \; [A^{*}_{\w \w_2} f^{\mathscr{I}^{+}\;*}_{\w_2}(x')  + B^{*}_{\w \w_2} f^{\mathscr{I}^{+}}_{\w_2}(x')]\Big] \right.  \nonumber \\
 & & \left. + \Big[f^{H^{+}}_\w(x') +  \int_0^\infty d\w_1 \; [A_{\w \w_1} f^{\mathscr{I}^{+}}_{\w_1}(x')  + B_{\w \w_1} f^{\mathscr{I}^{+}\;*}_{\w_1})(x')]\Big] \right.  \nonumber \\
 & & \Big[f^{H^{+} \;*}_\w(x) +  \int_0^\infty d\w_2 \; [A^{*}_{\w \w_2} f^{\mathscr{I}^{+}\;*}_{\w_2}(x)  + B^{*}_{\w \w_2} f^{\mathscr{I}^{+}}_{\w_2}(x)]\Big] \Big\}\;. \label{G1-1} \eea
  Expanding the integrand of the integral over $\w$ results in three types of expressions: an integral consisting of products of the modes $f^{H^+}_\w$, which we  call $G_A$, another integral which includes cross products between the modes $f^{H^+}_\w$ and  $f^{\mathscr{I}^+}_\w$, which we call $G_B$, and finally an integral consisting of products of the modes  $f^{\mathscr{I}^+}_\w$, which we called $G_C$\;.

 To renormalize we follow a procedure equivalent to that outlined in Sec.~\ref{sec-4D-method}.
 We begin by subtracting off the integrals with the integrand evaluated in the large $\w$ limit.  When we add them back, we get contributions that are identical to those obtained for the Unruh state.  We are not quite subtracting off the Unruh modes because the large $\w$ solutions have a dependence on $v_H$.  However, when the subtracted terms are added back and the integral over $\w$ is computed, then factors of $\delta(\w_1 - \w_2)$ and $\delta(\w_1 + \w_2)$ are obtained.  Note that terms proportional to $\delta(\w_1 + \w_2)$ vanish.  For the ones that do not vanish, once one integrates over say $\w_2$,  the dependence on $v_H$ vanishes.

 In Appendix~\ref{appendix-B} it is shown that when this method is applied to $ G^{(1)}(x,x')$, the $\Delta G_A$ term vanishes.  It is also shown that, while the $\Delta G_B$ term does not vanish, it does not contribute to the stress-energy tensor.  As a result, the only term that contributes to $\Delta \la T_{ab} \ra$ is $\Delta G_C(x,x')$ which has the form
\bea
\Delta G_C(x,x')&=& \int_0^{\infty} d\w_1 \int_0^{\infty} d\w_2 \Big\{ [f_{\w_1}^{\mathscr{I^+}}(x)f_{\w_2}^{\mathscr{I^+}\; *}(x')+f_{\w_1}^{\mathscr{I^+}}(x')f_{\w_2}^{\mathscr{I^+}\;*}(x)]\Delta I_1 \nonumber \\ & &
+\; [f_{\w_1}^{\mathscr{I^+}}(x)f_{\w_2}^{\mathscr{I^+}}(x')+f_{\w_1}^{\mathscr{I^+}}(x')f_{\w_2}^{\mathscr{I^+}}(x)]\Delta I_2 \nonumber \\ & &
+\;[f_{\w_1}^{\mathscr{I^+}\;*}(x)f_{\w_2}^{\mathscr{I^+}\;*}(x')+f_{\w_1}^{\mathscr{I^+}\;*}(x')f_{\w_2}^{\mathscr{I^+}\;*}(x)]\Delta I_3 \nonumber \\ & &
+\;[f_{\w_1}^{\mathscr{I^+}\;*}(x)f_{\w_2}^{\mathscr{I^+}}(x')+f_{\w_1}^{\mathscr{I^+}\; *}(x')f_{\w_2}^{\mathscr{I^+}}(x)]\Delta I_4 \Big\} \;, \label{Del-GC}
\eea
with
\bes \bea
\Delta I_1 &=& \int _0^{\infty} d\w \left\{A_{\w \w _1}A_{\w \w _2}^{*}-\mathcal{O}(A_{\w \w _1}A_{\w \w _2}^{*})\right\} \label{Del-I1}\;,\\
\Delta I_2 &=& \int _0^{\infty} d\w \left\{ A_{\w \w _1}B_{\w \w _2}^{*}-\mathcal{O}(A_{\w \w _1}B_{\w \w _2}^{*})\right\} \label{Del-I2}\;,\\
\Delta I_3 &=& \int _0^{\infty} d\w \left\{B_{\w \w _1}A_{\w \w _2}^{*}-\mathcal{O}(B_{\w \w _1}A_{\w \w _2}^{*})\right\}\label{Del-I3}\;,\\
\Delta I_4 &=& \int _0^{\infty} d\w \left\{B_{\w \w _1}B_{\w \w _2}^{*}-\mathcal{O}(B_{\w \w _1}B_{\w \w _2}^{*})\right\}\; \label{Del-I4}.
\eea  \label{In}  \ees
Here $\mathcal{O}$ indicates the asymptotic behavior of the matching coefficients for $\w \gg \w_{1,2}$.

The integrals in~\eqref{In} can be computed analytically. Substituting the explicit expression for A from~\eqref{AI2} into~\eqref{Del-I1} gives
\bes \bea  \Delta I_{1} &=&  \frac{1}{4 \pi^2  \sqrt{ \w_1 \w_2}} (4 M)^{i 4 M (\w_1-\w_2)} e^{i v_H(\w_1 - \w_2)} \Gamma(1-i 4M\w_1) \Gamma(1+i4M\w_2)
  \Delta K_1 \;, \label{Del-I1-1} \\
  \Delta K_{1} &=&   \lim_{\Lambda \to \infty} (-i)^{i 4 M \w_1} (i)^{-i 4 M \w_2}\bigg\{ \left[ \int_{0}^\Lambda d\w \frac{\w}{ ( \w-\w_1 + i \epsilon_1)^{1-i4M\w_1} ( \w-\w_2 - i \epsilon_2)^{1+i4M\w_2}}  \right. \nonumber \\   && \left. \qquad
- \int_{1}^\Lambda d\w \,\w^{-1+i4M(\w_1-\w_2)} \right]- \int_0^1  d\w \, \w^{-1+i4M(\w_1-\w_2)} \bigg\} \;. \label{DelK1-1}
\eea \ees
First we compute the indefinite integrals and evaluate them at the limits. Since $\epsilon_1$ and $\epsilon_2$  go to $O^+$ at the end of the calculation, it is acceptable to add terms containing them to the exponents. The first indefinite integral is
 \bea \Delta K_{1a} &=& (-i)^{i 4 M \w_1} (i)^{-i 4 M \w_2} \int_{0}^\Lambda d\w \frac{\w}{ ( \w-\w_1 + i \epsilon_1)^{1-i4M(\w_1- i \epsilon_1)} ( \w-\w_2 - i \epsilon_2)^{1+i4M(\w_2+i \epsilon_2)}}\nonumber \\
                        &=&  (-i)^{i 4 M \w_1} (i)^{-i 4 M \w_2} \left[-i \frac{(\w-\w_1+i \epsilon_1)^{-i4M(\w_1- i \epsilon_1)} (\w-\w_2-i \epsilon_2)^{-i 4M(\w_2+ i \epsilon_2)}}{4 M (\w_1-\w_2 - i \epsilon_1 - i \epsilon_2)} \right]_0^\Lambda  \nonumber \\
                        &=& -i (-i)^{i 4 M \w_1} (i)^{-i 4 M \w_2} \left[ \frac{(\Lambda-\w_1)^{i4M(\w_1)} (\Lambda-\w_2)^{-i 4M(\w_2)}}{4 M (\w_1-\w_2) - i \epsilon_1 - i \epsilon_2} \right.
                         \nonumber \\
                         & & \left. \; +  i \frac{(-\w_1)^{i4M(\w_1)} (-\w_2)^{-i 4M(\w_2)}}{4 M (\w_1-\w_2 - i \epsilon_1 - i \epsilon_2)} \right] \;.  \eea
Note that after evaluating the integral at the limits, $\epsilon_1$ and $\epsilon_2$ are set to zero in the exponents because they have no effect there.
Also, each term is a combination of a principle value and a term proportional to $\delta(\w_1-\w_2)$, thus
\bes \bea \Delta K_{1a} &=&   e^{2 \pi M (\w_1+\w_2)} \left[-i \frac{(\Lambda-\w_1)^{i4M\w_1} (\Lambda-\w_2)^{-i 4M\w_2}}{4 M (\w_1-\w_2)} + \frac{\pi}{4M} \delta(\w_1-\w_2) \right]
   \nonumber \\
 & &  +  e^{-2 \pi M (\w_1+\w_2)} \left[ i \frac{\w_1^{i4M\w_1} \w_2^{-i 4M\w_2}}{4 M (\w_1-\w_2)}  - \frac{\pi}{4M} \delta(\w_1-\w_2) \right] \;.  \label{K1a}
    \eea
Here we adopt the notation that the principle value of a term such as $\frac{1}{a \pm i \epsilon}$ is written as $\frac{1}{a}$.
   The second and third integrals in~\eqref{DelK1-1} are
  \bea \Delta K_{1b} &=& -(-i)^{i 4 M \w_1} (i)^{-i 4 M \w_2} \int_{1}^\Lambda d\w \w^{-1+i4M(\w_1-\w_2)} \nonumber \\
     &=&      i \frac{e^{2 \pi M(\w_1+\w_2)}}{4 M(\w_1-\w_2)} \left[ \Lambda^{i 4M(\w_1-\w_2)} - 1\right]  \;, \label{Del-K1b} \\
           \Delta K_{1c} &=&  -(-i)^{i 4 M \w_1} (i)^{-i 4 M \w_2} \int_0^1  d\w \, \w^{-1+i4M(\w_1-\w_2)} \nonumber \\
              &=&   - e^{2 \pi M(\w_1+\w_2)} \int_{-\infty}^0 dz e^{[i 4M (\w_1-\w_2)+ \epsilon] z}  =  -  \frac{e^{2 \pi M(\w_1+\w_2)}}{i 4M (\w_1-\w_2)+ \epsilon}  =  \frac{i e^{2 \pi M(\w_1+\w_2)}}{4M (\w_1-\w_2) -i \epsilon} \nonumber \\  & & \qquad =   \frac{ie^{2 \pi M(\w_1+\w_2)}}{4M (\w_1-\w_2)} -e^{2 \pi M(\w_1+\w_2)} \frac{\pi}{4M} \delta(\w_1-\w_2) \;, \label{Del-K1c}
\eea \ees
where in the integral for $\Delta K_{1c}$ the change of variable $z = \log \w$ has been made and an integrating factor $\epsilon$ has been inserted.
Combining these results, one finds
\bes \bea \Delta K_1 &=&    e^{-2 \pi M (\w_1+\w_2)} \left[ i \frac{\w_1^{i4M\w_1} \w_2^{-i 4M\w_2}}{4 M (\w_1-\w_2)}  - \frac{\pi}{4M} \delta(\w_1-\w_2) \right] \nonumber \\
   & =&  i  e^{-2 \pi M (\w_1+\w_2)} \frac{\w_1^{i4M\w_1} \w_2^{-i 4M\w_2}}{4 M (\w_1-\w_2 - i \epsilon)} \;. \label{Del-K1-1} \eea
Substituting ~\eqref{Del-K1-1} into~\eqref{Del-I1-1} gives
\bea \Delta I_1 &=& \frac{i}{4 \pi^2  \sqrt{ \w_1 \w_2}} (4 M)^{i 4 M (\w_1-\w_2)} e^{i v_H(\w_1 - \w_2)}   \Gamma(1-i 4M\w_1) \Gamma(1+i4M\w_2) \nonumber \\
  & & \qquad \times \, e^{-2 \pi M( \w_1+\w_2)} \frac{w_1^{i4M\w_1} \w_2^{-i 4M\w_2}}{4 M (\w_1-\w_2 - i \epsilon)} \;. \label{Del-I1-2}
       \eea \ees
Note that this is a finite contribution to $\Delta G_c$ because of the factor of $e^{-2 \pi M( \w_1+\w_2)}$.

Next consider  $\Delta I_4$ which is the other term with non-vanishing delta functions.
\bes \bea  \Delta I_{4} &=&  \frac{1}{4 \pi^2  \sqrt{ \w_1 \w_2}} (4 M)^{-i 4 M (\w_1-\w_2)} e^{-i v_H(\w_1 - \w_2)} \Gamma(1+i 4M\w_1) \Gamma(1-i4M\w_2)
  \Delta K_4 \;, \label{Del-I4-1} \\
  \Delta K_{4} &=&   \lim_{\Lambda \to \infty} (-i)^{-i 4 M \w_1} (i)^{i 4 M \w_2}\bigg\{ \left[ \int_{0}^\Lambda d\w \frac{\w}{ ( \w+\w_1 + i \epsilon_1)^{1+i4M\w_1} ( \w+\w_2 - i \epsilon_2)^{1-i4M\w_2}}  \right. \nonumber \\   && \left. \qquad
- \int_{1}^\Lambda d\w \,\w^{-1-i4M(\w_1-\w_2)} \right]- \int_0^1  d\w \, \w^{-1-i4M(\w_1-\w_2)} \bigg\} \;. \label{DelK4-1} \;
\eea \ees
The integrals in $ \Delta K_{4}$ can be computed analytically
\bea  \Delta K_{4a} &=& e^{-2\pi M(\w_1+\w_2)} \int_{0}^{\Lambda} d\w  \frac{\w}{( \w+\w_1+i \epsilon_1)^{1+i4M(\w_1+i \epsilon_1)} (\w+\w_2-i \epsilon_2)^{1-i4M(\w_2-i \epsilon_2)}} \nonumber \\
              &=& e^{-2\pi M(\w_1+\w_2)} \frac{i}{4 M} \left[ \frac{(\Lambda+\w_1)^{-i4M\w_1} (\Lambda+\w_2)^{i4M\w_2}}{\w_1-\w_2+ i (\epsilon_1+\epsilon_2)}
               -  \frac{\w_1^{-i4M\w_1} \w_2^{i4M\w_2}}{\w_1-\w_2+ i (\epsilon_1+\epsilon_2)}   \right] \nonumber \\
              \Delta K_{4b}& =&  e^{-2\pi M(\w_1+\w_2)} \left[ -\frac{\Lambda^{-i 4 M  (\w_1-\w_2)}}{-i 4M(\w_1-\w_2)} + \frac{1}{-i 4M(\w_1-\w_2)} \right] \,, \nonumber \\
              \Delta K_{4c} &= &  - e^{-2\pi M(\w_1+\w_2)} \int_{-\infty}^0 dz \, e^{[-i 4M(\w_1-\w_2)+\epsilon]z} = -\frac{ e^{-2\pi M(\w_1+\w_2)}}{-i 4M(\w_1-\w_2)+\epsilon}
             \nonumber \\
              &=& -i \frac{e^{-2\pi M(\w_1+\w_2)}}{4 M (\w_1-\w_2) + i \epsilon}  = -i  \frac{e^{-2\pi M(\w_1+\w_2)}}{4 M (\w_1-\w_2)} - e^{-2\pi M(\w_1+\w_2)} \frac{\pi}{4M}   \delta(\w_1-\w_2)  \;.
                \eea
Both terms in $\Delta K_{4a}$ can be written in terms of their principle values added to a term proportional to $\delta(\w_1-\w_2)$. Combining these terms, the following expression for $\Delta K_4$ is obtained
\bes \bea  \Delta K_4 &=&   e^{-2\pi M(\w_1+\w_2)}\left[-i \frac{\w_1^{-i4M\w_1} \w_2^{i4M\w_2}}{4M(\w_1-\w_2)} - \frac{ \pi }{4M}   \delta(\w_1-\w_2)\right]
   \nonumber \\
     &=& - i e^{-2\pi M(\w_1+\w_2)} \; \frac{\w_1^{-i4M\w_1} \w_2^{i4M\w_2}}{4M(\w_1-\w_2 + i \epsilon)}  \;.
     \eea
Finally
\bea \Delta I_4 &=& -\frac{i}{4 \pi^2  \sqrt{ \w_1 \w_2}} (4 M)^{-i 4 M (\w_1-\w_2)} e^{-i v_H(\w_1 - \w_2)}   \Gamma(1+i 4M\w_1) \Gamma(1-i4M\w_2) \nonumber \\
    & & \qquad \times\, e^{-2\pi M(\w_1+\w_2)} \; \frac{\w_1^{-i4M\w_1} \w_2^{i4M\w_2}}{4M(\w_1-\w_2 + i \epsilon)}  \;. \label{Del-I4-2} \eea \ees
Note that if we let $\w_1 \leftrightarrow \w_2$ in the expression~\eqref{Del-I1-2} for $\Delta I_1$, then we get $\Delta I_4$ in~\eqref{Del-I4-2}.
It is also true that if this switch is made in the entire contribution to the two-point function from $\Delta I_1$ then that is equal to the contribution from $\Delta I_4$.
Next consider $\Delta I_2$
\bes \bea  \Delta I_{2} &=&  -\frac{1}{4 \pi^2  \sqrt{ \w_1 \w_2}} (4 M)^{i 4 M (\w_1+\w_2)} e^{i v_H(\w_1 + \w_2)} \Gamma(1-i 4M\w_1) \Gamma(1-i4M\w_2)
  \Delta K_2 \;, \label{Del-I2-1} \\
  \Delta K_{2} &=&   \lim_{\Lambda \to \infty} (-i)^{i 4 M \w_1} (i)^{i 4 M \w_2}\bigg\{ \left[ \int_{0}^\Lambda d\w \frac{\w}{ ( \w-\w_1 + i \epsilon_1)^{1-i4M\w_1} ( \w+\w_2 - i \epsilon_2)^{1-i4M\w_2}}  \right. \nonumber \\   && \left. \qquad
- \int_{1}^\Lambda d\w \,\w^{-1+i4M(\w_1+\w_2)} \right]- \int_0^1  d\w \, \w^{-1+i4M(\w_1+\w_2)} \bigg\} \;, \label{DelK2-1}
\eea \ees
where the integrals in $ \Delta K_{2}$ can be computed analytically
\bes \bea  \Delta K_{2a} &=&  (-i)^{i 4 M \w_1} (i)^{i 4 M \w_2} \int_{0}^\Lambda d\w \frac{\w}{( \w-\w_1 + i \epsilon_1)^{1-i4M(\w_1-i \epsilon_1)} (\w+\w_2-i \epsilon_2)^{1-i4M(\w_2-i \epsilon_2)}}  \nonumber \\
   &=& -\frac{i}{4M} (-i)^{i 4 M \w_1} (i)^{i 4 M \w_2} \left\{ \frac{(\Lambda-\w_1)^{i4M \w_1} (\Lambda+\w_2)^{i4M\w_2}}{[\w_1+\w_2 - i(\epsilon_1 + \epsilon_2)]} \right. \nonumber \\  & & \left. -  \frac{(-\w_1)^{i4M\w_1} \w_2^{i4M\w_2-}}{[\w_1+\w_2 - i(\epsilon_1 + \epsilon_2)]} \right\} \nonumber \\
   &=& -\frac{i}{4M} e^{2 \pi M(\w_1 -\w_2)} \frac{(\Lambda-\w_1)^{i4M \w_1} (\Lambda+\w_2)^{i4M\w_2}}{[\w_1+\w_2 - i(\epsilon_1 + \epsilon_2)]}  \nonumber \\
     & &   + \frac{i}{4M} e^{-2 \pi M(\w_1+\w_2)} \frac{\w_1^{i4M\w_1} \w_2^{i4M\w_2}}{[\w_1+\w_2 - i(\epsilon_1 + \epsilon_2)]}\;,  \label{Del-K2a}  \\
  \Delta K_{2b} &=&  \frac{i}{4 M} e^{2 \pi M(\w_1 -\w_2)} \frac{\Lambda^{i 4M (\w_1+\w_2)}}{\w_1+\w_2}  - \frac{i}{4 M} e^{2 \pi M(\w_1 -\w_2)} \frac{1}{\w_1+\w_2} \;, \\
  \Delta K_{2c} &=&  - e^{2\pi M(\w_1-\w_2)} \int_{-\infty}^0 dz \, e^{[i 4M(\w_1+\w_2)+\epsilon]z} = -\frac{ e^{2\pi M(\w_1-\w_2)}}{i 4M(\w_1+\w_2)+\epsilon}
             \nonumber \\
              &=& i \frac{e^{2\pi M(\w_1-\w_2)}}{4 M (\w_1+\w_2) - i \epsilon}  = i  \frac{e^{2\pi M(\w_1-\w_2)}}{4 M (\w_1+\w_2)} - \frac{\pi}{4M}   \delta(\w_1+\w_2)  \;.
\eea \ees
Given that $\delta(\w_1+\w_2) = 0$ since the frequencies are all non-negative, one can set $\epsilon_1=\epsilon_2=0$. Then
\bes \be \Delta K_2 = i e^{-2 \pi M(\w_1+\w_2)} \frac{\w_1^{i4M\w_1} \w_2^{i4M\w_2}}{4M(\w_1+\w_2)} \;, \ee
and
\bea \Delta I_2 &=& -\frac{i}{4 \pi^2  \sqrt{ \w_1 \w_2}} (4 M)^{i 4 M (\w_1+\w_2)} e^{i v_H(\w_1 + \w_2)}  \Gamma(1-i 4M\w_1) \Gamma(1-i4M\w_2) \nonumber \\
   & & \times e^{-2 \pi M(\w_1+\w_2)} \frac{\w_1^{i4M\w_1} \w_2^{i4M\w_2}}{4M(\w_1+\w_2)}  \;. \label{Del-I2-2} \eea  \ees

Comparing $\Delta I_2$ in~\eqref{Del-I2} and $\Delta I_3$ in~\eqref{Del-I3}, one can immediately see that their contributions to the two-point function,~\eqref{Del-GC}, are the complex conjugate of each other if one also takes $\w_1 \leftrightarrow \w_2$ in the contribution from $\Delta I_2$
 \bea \Delta I_3 &=&  (\Delta I_2)^{*}\nonumber\\
 &=&\frac{i}{4 \pi^2  \sqrt{ \w_1 \w_2}} (4 M)^{-i 4 M (\w_1+\w_2)} e^{-i v_H(\w_1 + \w_2)}  \Gamma(1+i 4M\w_1) \Gamma(1+i4M\w_2) \nonumber \\
   & & \times e^{-2 \pi M(\w_1+\w_2)} \frac{\w_1^{-i4M\w_1} \w_2^{-i4M\w_2}}{4M(\w_1+\w_2)} \;.  \label{Del-I3-2} \eea

Substituting~\eqref{Del-I1-2},~\eqref{Del-I4-2},~\eqref{Del-I2-2}, and~\eqref{Del-I3-2} into~\eqref{Del-GC} one finds
\bea
\Delta G_C(x,x')&=&\mathfrak{R}\Bigg\{ \frac{i}{8\pi^3}\int_0^{\infty}\frac{d\w_1}{\w_1}\int_0^{\infty}\frac{d\w_2}{\w_2}e^{-2\pi M(\w_1+\w_2)} \nonumber \\ & &
\times \bigg\{ \Big[e^{-i\w_1 u_s+i\w_2 u'_s}+e^{-i\w_1 u'_s+i\w_2 u_s}\Big] \frac{(4M\w_1e^{\frac{v_H}{4M}})^{4iM\w_1}}{(4M\w_2e^{\frac{v_H}{4M}})^{4iM\w_2}}\frac{\Gamma(1-4iM\w_1)\Gamma(1+4iM\w_2)}{4M(\w_1-\w_2-i\epsilon)} \nonumber \\ & &
-\Big[e^{-i\w_1 u_s-i\w_2 u'_s}+e^{-i\w_1 u'_s-i\w_2 u_s}\Big](4M\w_1e^{\frac{v_H}{4M}})^{4iM\w_1}(4M\w_2e^{\frac{v_H}{4M}})^{4iM\w_2} \nonumber \\ & &
 \times \frac{\Gamma(1-4iM\w_1)\Gamma(1-4iM\w_2)}{4M(\w_1+\w_2)}\bigg\} \Bigg\}\;. \label{Delta-GC-fin}
 \eea
There are infrared divergences in this expression.  However, it is easy to see that the derivatives in the general formula for the stress-energy tensor~\eqref{Tab-class} bring down factors of $\w_1$ and $\w_2$ which remove these infrared divergences.  Recalling that $\Delta G_C$ is the only contribution to $\la T_{ab} \ra$ from  $\Delta G^{(1)}$, it is straight-forward to show using~\eqref{Delta-Tab-2},~\eqref{T_trB},~\eqref{rtpttp},~\eqref{DTrt-DTtt}, and~\eqref{Delta-GC-fin} that
\bea
\Delta \langle T_{tt} \rangle&=&  -(1-\frac{2M}{r}) \lim_{x' \to x} \frac{1}{4}(\Delta G_{C\; ;t';r}+\Delta G_{C \;;t;r'})\nonumber \\
&=&\mathfrak{R}\Bigg\{ \frac{i}{8\pi^3}\int_0^{\infty}d\w_1 \; \int_0^{\infty}d\w_2 \; e^{-2\pi M(\w_1+\w_2)} \nonumber \\ & &
\times \bigg\{ e^{i(\w_2-\w_1) u_s} \frac{(4M\w_1e^{\frac{v_H}{4M}})^{4iM\w_1}}{(4M\w_2e^{\frac{v_H}{4M}})^{4iM\w_2}}\frac{\Gamma(1-4iM\w_1)\Gamma(1+4iM\w_2)}{4M(\w_1-\w_2-i\epsilon)} \nonumber \\ & &
+ e^{-i(\w_2 +\w_1)u_s}(4M\w_1e^{\frac{v_H}{4M}})^{4iM\w_1}(4M\w_2e^{\frac{v_H}{4M}})^{4iM\w_2} \nonumber \\ & &
 \times \frac{\Gamma(1-4iM\w_1)\Gamma(1-4iM\w_2)}{4M(\w_1+\w_2)}\bigg\} \Bigg\}\;.
\label{2D-Ttt} \eea

 The integral over $\w_2$ of the first term inside the curly bracket can be written in the form
  \bea
 \Delta \langle T_{tt} \rangle _1 &=& \int_0^{\infty} d\w_2 \; \frac{f(\w_2)}{\w_1-\w_2-i\epsilon}=  \int_0^\infty d\w_2 \;\left[ \frac{f(\w_2)}{\w_1-\w_2}+i\pi\delta(\w_1-\w_2) \right] \nonumber \\
 &=& \lim_{\epsilon \to 0^+} \left[ \int_0^{\w_1-\epsilon} d \w_2  \frac{f(\w_2)}{\w_1-\w_2} + \int_{\w_1+\epsilon}^\infty d \w_2  \frac{f(\w_2)}{\w_1-\w_2} \right] + i \pi f(\w_1)
\;, \label{CPV}
  \eea
where the definition of the Cauchy principal value integral has been explicitly used.

Thus, extracting the explicit form of the $f(\w_2)$  from ~\eqref{2D-Ttt} and substituting it into ~\eqref{CPV} yields
\bea
\Delta \langle T_{tt} \rangle _1
 &=&\mathfrak{R}\Bigg\{ \frac{i}{8\pi^3}\int_0^{\infty} d\w_1\; \Big[\int_0^{\infty} d\w_2 \;  e^{-2\pi M(\w_1+\w_2)} e^{i(\w_2- \w_1) u_s} \nonumber \\ & & \times
 \frac{(4M\w_1e^{\frac{v_H}{4M}})^{4iM\w_1}}{(4M\w_2e^{\frac{v_H}{4M}})^{4iM\w_2}}\frac{\Gamma(1-4iM\w_1)\Gamma(1+4iM\w_2)}{4M(\w_1-\w_2)}\Big]\Bigg\}+
 \frac{1}{4\pi^2}\int_0^{\infty} d\w_1 \; \frac{e^{-4\pi M\w_1}}{4M} \nonumber \\ & &
\times \Gamma(1-4iM\w_1)\Gamma(1+4iM\w_1)\;.  \label{Del-Ttt}
 \eea

The stress-energy tensor for a massless minimally coupled scalar field in the 2D collapsing null shell spacetime has been previously computed analytically using a different method~\cite{hiscock}\cite{Fabbri:2005mw} and the stress-energy tensor for the Unruh state has also been computed analytically~\cite{Davies-Fulling-Unruh,Fabbri:2005mw}.  For the difference one finds
 \bea
\Delta \la T_{uu} \ra &=& -\frac{1}{24\pi}\left[\frac{8M}{(u-v_0)^3}+\frac{24M^2}{(u-v_0)^4}\right] - \frac{1}{768 \pi M^2}  \;, \nonumber  \\
\Delta \la T_{uv} \ra &=& \Delta \la T_{vv} \ra = 0 \;, \nonumber \\
\Delta \la T_{tt} \ra &=& \Delta \la T_{uu} \ra + 2 \Delta \la T_{uv} \ra + \Delta \la T_{vv} \ra \nonumber \\
 & = & -\frac{1}{24\pi}\left[\frac{8M}{(u-v_0)^3}+\frac{24M^2}{(u-v_0)^4}\right]  - \frac{1}{768 \pi M^2}  \;.
\label{Hiscock-Result} \eea

\begin{figure}[h]
\centering
\includegraphics [totalheight=0.45\textheight]{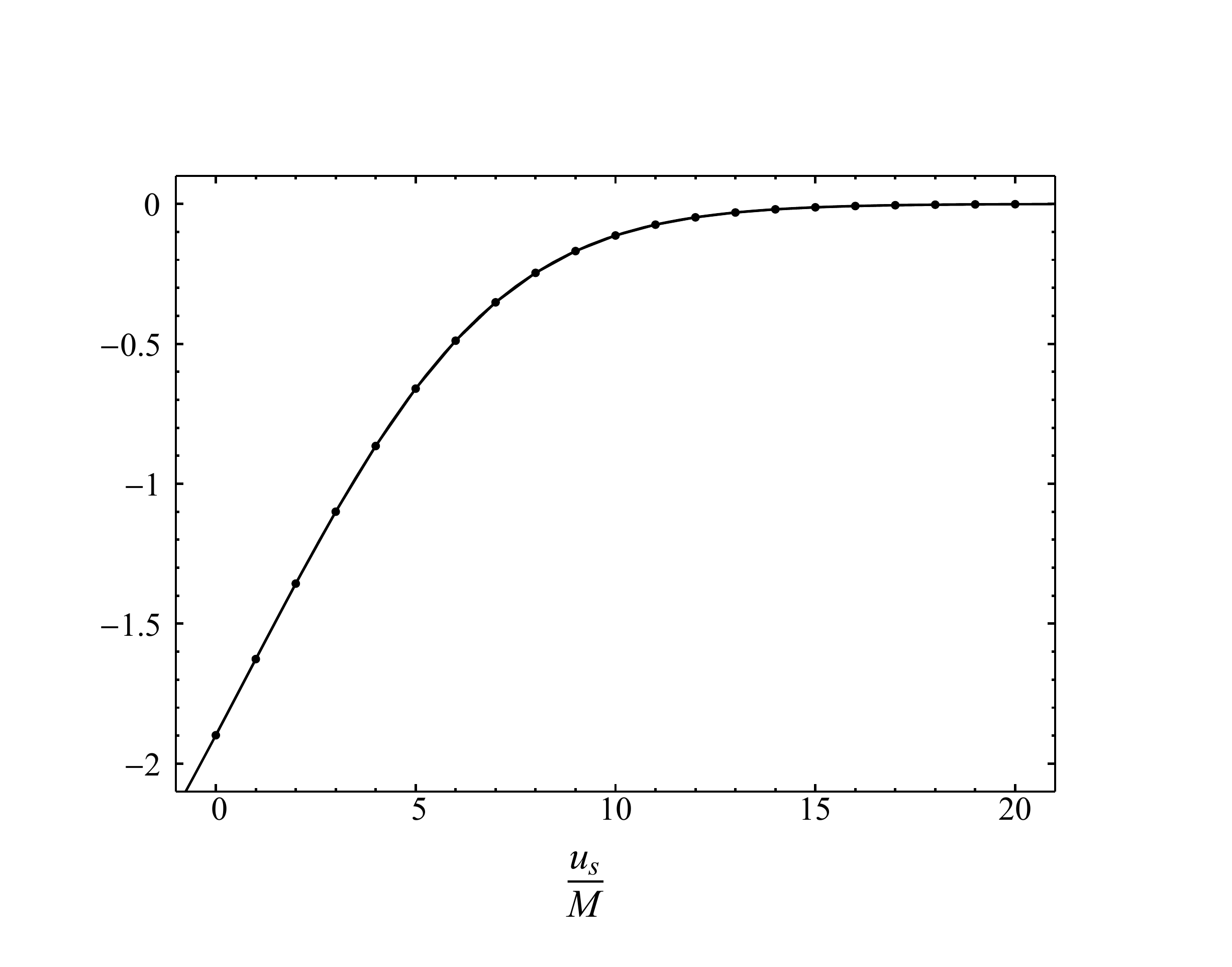}
\caption{The quantity $10^{4}M^2\Delta \langle T_{tt} \rangle$ is plotted for the massless minimally coupled scalar field in the region exterior to the null shell
and to the event horizon. The dots correspond to the results of the numerical computations.  The solid curve represents the analytic results in~\eqref{Hiscock-Result}.   }
\label{fig:Ttt}
\end{figure}

Both terms in~\eqref{Del-Ttt} have been computed numerically. In the first integral, the numerical computation has been performed  by the symmetric removal of the neighborhood with radius $\epsilon$ about the singular points of the integrand, $\w_1=\w_2$. The integral of the second term in ~\eqref{2D-Ttt}  has also been found by a more straightforward numerical method.
 Our results for $\Delta \langle  T_{tt} \rangle$ in~\eqref{2D-Ttt} are shown in Figure~\ref{fig:Ttt}. Although it is not possible to detect this from the plot, our numerical results agree with the analytical results in~\cite{hiscock} \cite{Fabbri:2005mw} to more than ten digits.

It is worth mentioning that in 2D, once  $\Delta \langle T_{tt} \rangle$ is numerically computed,  $\Delta \langle T_{rr} \rangle$ and  $\Delta \langle T_{tr} \rangle$ can be easily determined from the relations~\eqref{T_rrB} and~\eqref{DTrt-DTtt}.

\section{Summary}

We have presented a method of numerically computing the stress-energy tensor for a massless minimally coupled scalar field in the case when a black hole that is formed from the
collapse of a spherically symmetric null shell in four dimensions.  There are two primary parts to the method.  The first is to expand the mode functions in the natural {\it in} vacuum state in terms of a complete set of mode functions in the part of Schwarzschild spacetime that is outside of the event horizon of the black hole. Expressions have been found for the matching coefficients that involve integrals of these mode functions over the trajectory of the null shell.

The second part of the method involves subtracting the unrenormalized expression for the stress-energy tensor in the Unruh state from the expression for the unrenormalized stress-energy tensor in the {\it in} vacuum state.  Since the ultraviolet divergences in the stress-energy tensor are independent of the state, this difference is finite.  Then one can add to this the renormalized expression for the stress-energy tensor in the Unruh state that has already been computed~\cite{levi-et-al-kerr} and the result is the full renormalized stress-energy tensor for the {\it in} vacuum state.

We have tested the first part of the method by  analytically  computing the matching coefficients in the 2-D case and reconstructing the mode functions for the {\it in} vacuum state.  We have also analytically computed the matching coefficients in 4D for the spherically symmetric mode functions (those with $\ell = 0$) in the {\it in} vacuum state for a simple model in which the effective potential in the mode equation is proportional to a Dirac delta function.  In this case it was possible to analytically compute the part of the mode function in the {\it in} vacuum state that is proportional to $e^{-i \w v}$ inside the null shell and to verify that it gives the known result on the matching surface.  Finally, for the actual case of a collapsing null shell in 4D, we have analytically computed parts of the matching coefficients and used those parts to numerically compute part of one of the {\it in} modes on the future horizon and shown that it has the correct value at the point where the future horizon intersects the null shell trajectory.
 We also used the method to numerically compute the stress-energy tensor for the 2-D case and found that the result was in excellent agreement with the known analytic result.

These tests provide substantial evidence that the method will work and that it will be possible to numerically compute the exact renormalized stress-energy tensor for a massless minimally coupled scalar field in a 4D spacetime in which a black hole forms from the collapse of a spherically symmetric null shell.  Work on that computation is in progress.

\acknowledgments

P. R. A. would like to thank Eric Carlson, Charles Evans, Adam Levi, and Amos Ori for helpful conversations and Adam Levi for sharing some of his numerical data.
A.F. acknowledges partial financial support from the
Spanish Ministerio de Ciencia e Innovaci\'on grant
FIS2017-84440-C2-1-P and from the Generalitat Valenciana grant
PROMETEO/2020/079.  This work was supported in part by the National Science Foundation under Grants No.  PHY-1308325, PHY-1505875, and PHY-1912584 to Wake Forest University.
Some of the numerical work was done using the WFU DEAC cluster; we thank the WFU Provost's Office and Information
Systems Department for their generous support.

\appendix
\section{Lagrange Inversion Theorem applied to the Lambert W function}
\label{appendix-A}

In~\cite{Corless} the relation
\be e^{-c W(x)} =  \sum_{n=0}^\infty \frac {c(n+c)^{n-1}}{n!} (-x)^n \;. \label{final-Lambert-1} \ee
is derived for any complex constant $c$.  An alternative derivation is given here.  It is based on the Lagrange inversion theorem~\cite{Jacobi}.  In~\cite{Gessel} different forms for the Lagrange inversion theorem are given, one of which we use here. To state the form that is most useful to us we use the notation in~\cite{Gessel} that
if $f(x)$ is expanded in a Laurent series then $[x^n] f(x)$ denotes the coefficient of $x^n$ in that series.
Then a statement of the theorem is:
Suppose $f$ is a function of $x$ and there is a relation of the following form
\bea f(x) = x R(f(x)) \;, \label{f-relation} \eea
where $R(t)$ is a power series in $t$.  Suppose further that $\phi(t)$ is also a Laurent series in $t$.  Then for any nonzero integer $n$, $\phi(f(x))$ can be expressed in terms of a unique power series in $x$ with coefficients
\bea
[x^n]\phi(f)\equiv \frac{1}{n!} \frac{d^{n} \phi(f(x))}{dx^n} = \frac{1}{n}[t^{n-1}] \phi'(t)R(t)^n\;, \label{LIT}
\eea
where the interpretation of the far right hand side is that one first expands the function $\frac{d\phi(t)}{dt} R(t)$ in powers of $t$, then chooses the coefficient of the term proportional to $t^{n-1}$ in that series and divides that coefficient by $n$.

To use this to obtain a power series for the function $e^{-c W(x)}$, note that the Lambert W function satisfies the relation
\be W(x) = x e^{-W(x)}  = x \sum_{n=0}^\infty \frac{(-W(x))^n}{n!}  \;. \label{Lambert-2} \ee
Thus we can choose the function $R(t)$ in~\eqref{f-relation} to be $R(t) = e^{-t}$.
We also choose $\phi(t) = e^{-c t}$.  Then
\bea [x^n] e^{-c W(x)} &=& \frac{1}{n} [t^{n-1}]\phi'(t) R^n(t)  \nonumber \\
                       &=& -\frac{c}{n} [t^{n-1}] e^{-(c+n) t} =  -\frac{c}{n!} [-(c+n)]^{n-1}  \nonumber \\
                       &=& (-1)^n \frac{c}{n!} (c +n)^{n-1}  \;. \eea
Equation~\eqref{final-Lambert-1} follows immediately from this.

\section{Contributions to the stress-energy tensor}
\label{appendix-B}

The calculations in this appendix are done entirely for the Schwarzschild geometry.  Therefore for simplicity we use $t$ and $u$ to denote the usual time coordinate and the right moving radial null coordinate in Schwarzschild spacetime.

In Sec.~\ref{sec:Tab-2D} it is mentioned that for the null shell spacetime in 2D the Hadamard Green's function in~\eqref{G1-1} can be broken into three parts.  One of these, which we call $G_A^{(1)}(x,x')$, includes $f^{H^+}$ and its complex conjugate and is given by the expression
\bea
 G^{(1)}_A(x,x')= \int_0^{\infty}d\w \Big\{f^{H^+}_{\w}(x)f^{H^+ \; *}_{\w}(x')+f^{H^+}_{\w}(x')f^{H^+ \; *}_{\w}(x)\Big\}
 \;. \label{GA}
\eea
For the Unruh state the corresponding contribution to $G^{(1)}(x,x')$ is exactly the same so $\Delta G^{(1)}_A(x,x') = 0$.

A second part, $G_B^{(1)}(x,x')$, has terms involving products of $f^{H^+}$ and its complex conjugate with $f^{\mathscr{I^+}}$ and its complex conjugate such that
 \bea
 G^{(1)}_{B}(x,x') &=& \int_0^\infty d \w \;\Big\{ \int_0^\infty d\w_2 \;   \Big[A^{*}_{\w \w_2} f^{H^{+}}_\w(x) f^{\mathscr{I}^{+}\;*}_{\w_2}(x')  + B^{*}_{\w \w_2}  f^{H^{+}}_\w(x)f^{\mathscr{I}^{+}}_{\w_2}(x')
\Big] \nonumber \\ &&
+  \int_0^\infty d\w_1 \; \Big[A_{\w \w_1}  f^{H^{+} \;*}_\w(x') f^{\mathscr{I}^{+}}_{\w_1}(x)  + B_{\w \w_1}  f^{H^{+} \;*}_\w(x') f^{\mathscr{I}^{+}\;*}_{\w_1}(x)\Big]\nonumber \\ &&
 \int_0^\infty d\w_2 \;\Big[A^{*}_{\w \w_2} f^{H^{+}}_\w(x')  f^{\mathscr{I}^{+}\;*}_{\w_2}(x)  + B^{*}_{\w \w_2} f^{H^{+}}_\w(x')  f^{\mathscr{I}^{+}}_{\w_2}(x)\Big] \nonumber \\ &&
 \; \int_0^\infty d\w_1 \;\Big[ A_{\w \w_1} f^{H^{+} \;*}_\w(x) f^{\mathscr{I}^{+}}_{\w_1}(x')  + B_{\w \w_1}  f^{H^{+} \;*}_\w(x)f^{\mathscr{I}^{+}\;*}_{\w_1})(x')\Big]\Big\}
\;. \label{GB} \eea
There is no contribution to $G^{(1)}(x,x')$ which has terms of this form if the field is in the Unruh state, so there is no subtraction term and $
\Delta G_B^{(1)}(x,x') = G_B^{(1)}(x,x')$.

While $ G^{(1)}_{B}(x,x')$ contributes to the two-point function, we next show that its contribution to the stress-energy tensor is zero.
Substituting~\eqref{f-form-2D} into~\eqref{GB} and using~\eqref{right-moving-2D} and~\eqref{left-moving-2D} , one readily finds that
 \bea
\Big[ G^{(1)}_{B}(x,x') \Big]_{;t;t'}&=&\frac{1}{4\pi}\int_0^\infty d \w \; \Big\{\int_0^\infty d\w_2 \;  \sqrt{\w\w_{2}} \; (A^{*}_{\w \w_2}e^{-i\w v+i\w_{2}u'} -B^{*}_{\w \w_2} e^{-i\w v-i\w_{2} u'}) \nonumber \\ &&
+ \int_0^\infty d\w_1 \;  \sqrt{\w\w_{1}} \;(A_{\w \w_1}e^{i\w v'-i\w_{1}u} -B_{\w \w_1} e^{i\w v'+i\w_{1} u}) \nonumber \\ &&
+ \int_0^\infty d\w_2 \;  \sqrt{\w\w_{2}} \; (A^{*}_{\w \w_2}e^{-i\w v'+i\w_{2}u} -B^{*}_{\w \w_2} e^{-i\w v'-i\w_{2} u}) \nonumber \\ &&
+ \int_0^\infty d\w_1 \;  \sqrt{\w\w_{1}} \; (A_{\w \w_1}e^{i\w v-i\w_{1}u'} -B_{\w \w_1} e^{i\w v+i\w_{1} u'})\Big\}
\;, \label{G_tt'} \eea

 \bea
\Big[ G^{(1)}_{B}(x,x') \Big]_{;r;r'}&=&\frac{1}{4\pi(1-\frac{2M}{r})^2} \int_0^\infty d \w \;\Big\{ \int_0^\infty d\w_2 \;  \sqrt{\w\w_{2}} \; (-A^{*}_{\w \w_2}e^{-i\w v+i\w_{2}u'} +B^{*}_{\w \w_2} e^{-i\w v-i\w_{2} u'}) \nonumber \\ &&
+ \int_0^\infty d\w_1 \;  \sqrt{\w\w_{1}} \; (-A_{\w \w_1}e^{i\w v'-i\w_{1}u} +B_{\w \w_1} e^{i\w v'+i\w_{1} u}) \nonumber \\ &&
+\int_0^\infty d\w_2 \;  \sqrt{\w\w_{2}} \; (-A^{*}_{\w \w_2}e^{-i\w v'+i\w_{2}u} +B^{*}_{\w \w_2} e^{-i\w v'-i\w_{2} u}) \nonumber \\ &&
+\int_0^\infty d\w_1 \;  \sqrt{\w\w_{1}} \; (-A_{\w \w_1}e^{i\w v-i\w_{1}u'}+B_{\w \w_1} e^{i\w v+i\w_{1} u'})\Big\}
\;. \label{G_rr'} \eea
From~\eqref{Delta-Tab-2} one finds
\bea
\langle T_{tt} \rangle=\frac{1}{4} \lim_{x \to x'} \Big[G_{;t;t'}+(1-\frac{2M}{r})^2G_{;r;r'}\Big] \label{T_ttB}
\; .\eea
By substituting ~\eqref{G_tt'} and ~\eqref{G_rr'}  into ~\eqref{T_ttB},  it is easy to see that the contribution to $\langle T_{tt}\rangle$ is zero.

Next consider the contribution of $ G^{(1)}_{B}(x,x')$ to $\langle T_{rr} \rangle$.
Using~\eqref{Delta-Tab-2} it is not hard to show that
\bea
\langle T_{rr} \rangle=\frac{1}{4} \lim_{x \to x'} \Big[\frac{G_{;t;t'}}{(1-\frac{2M}{r})^2}+G_{;rr'}\Big] \label{T_rrB}
\;. \eea
Together with ~\eqref{T_ttB}, one obtains
\bea
\langle T_{rr} \rangle=\frac {\langle T_{tt} \rangle}{\left(1-\frac{2M}{r}\right)^2}\;.
\eea
Thus $ G^{(1)}_{B}(x,x') $ does not contribute to  $\langle T_{rr}\rangle$ either.

Finally, we consider the contribution of  $ G^{(1)}_{B}(x,x') $  to $\langle T_{tr}\rangle$.  From~\eqref{Delta-Tab-2} one finds
\bea
\langle T_{tr} \rangle=\frac{1}{4} \lim_{x \to x'} \Big[G_{;t';r}+G_{;t;r'}\Big] \label{T_trB}
\;. \eea
Taking the derivative of ~\eqref{GB} with respect to $t$ and $r'$, one finds
 \bea
\Big[ G^{(1)}_{B}(x,x') \Big]_{;t;r'}&=&\frac{1}{4\pi(1-\frac{2M}{r'})} \int_0^\infty d \w \; \Big\{\int_0^\infty d\w_2 \;  \sqrt{\w\w_{2}} \; (-A^{*}_{\w \w_2}e^{-i\w v+i\w_{2}u'} +B^{*}_{\w \w_2} e^{-i\w v-i\w_{2} u'}) \nonumber \\ &&
+  \int_0^\infty d\w_1 \;  \sqrt{\w\w_{1}} \; (A_{\w \w_1}e^{i\w v'-i\w_{1}u} -B_{\w \w_1} e^{i\w v'+i\w_{1} u}) \nonumber \\ &&
+ \int_0^\infty d\w_2 \;  \sqrt{\w\w_{2}} \; (A^{*}_{\w \w_2}e^{-i\w v'+i\w_{2}u} -B^{*}_{\w \w_2} e^{-i\w v'-i\w_{2} u}) \nonumber \\ &&
+ \int_0^\infty d\w_1 \;  \sqrt{\w\w_{1}} \; (-A_{\w \w_1}e^{i\w v-i\w_{1}u'}+B_{\w \w_1} e^{i\w v+i\w_{1} u'})\Big\}
\;. \label{G_tr'} \eea
and taking the derivative of ~\eqref{GB} with respect to $t'$ and $r$ gives
 \bea
\Big[ G^{(1)}_{B}(x,x') \Big]_{;t';r}&=&\frac{1}{4\pi(1-\frac{2M}{r'})}\int_0^\infty d \w \;\Big\{ \int_0^\infty d\w_2 \;  \sqrt{\w\w_{2}} \; (A^{*}_{\w \w_2}e^{-i\w v+i\w_{2}u'} -B^{*}_{\w \w_2} e^{-i\w v-i\w_{2} u'}) \nonumber \\ &&
+ \int_0^\infty d\w_1 \;  \sqrt{\w\w_{1}} \; (-A_{\w \w_1}e^{i\w v'-i\w_{1}u} +B_{\w \w_1} e^{i\w v'+i\w_{1} u}) \nonumber \\ &&
+ \int_0^\infty d\w_2 \;  \sqrt{\w\w_{2}} \; (-A^{*}_{\w \w_2}e^{-i\w v'+i\w_{2}u} +B^{*}_{\w \w_2} e^{-i\w v'-i\w_{2} u}) \nonumber \\ &&
+\int_0^\infty d\w_1 \;  \sqrt{\w\w_{1}} \; (A_{\w \w_1}e^{i\w v-i\w_{1}u'}-B_{\w \w_1} e^{i\w v+i\w_{1} u'})\Big\}
\;. \label{G_t'r}
\eea
It is clear that $\Big[ G^{(1)}_{B}(x,x') \Big]_{;t';r} = - \Big[ G^{(1)}_{B}(x,x') \Big]_{;t;r'}$ and therefore that their contribution to  $\langle T_{tr}\rangle$ is zero.

The third part of $G^{(1)}(x,x')$ we call $G^{(1)}_C(x,x')$.  Its contribution to $\Delta \la T_{tt} \ra$ is given in Sec.~\ref{sec:Tab-2D}.

\section{Relation between two components of $\Delta \la T_{ab} \ra$}
\label{appendix-C}

The calculations in this appendix are done entirely for the Schwarzschild geometry.  Therefore for simplicity we use $t$ and $u$ to denote the usual time coordinate and the right moving radial null coordinate in Schwarzschild spacetime.

In this appendix a relation is derived between two components of $\Delta \la T_{ab} \ra$ in~\eqref{Delta-Tab-2} for the 2D collapsing null shell spacetime.
As shown in Appendix~\ref{appendix-B} only $\Delta G_C(x,x')$ in~\eqref{Del-GC} contributes to $\Delta \la T_{ab} \ra$.  The explicit form for $\Delta G_c(x,x')$ is
\bea
\Delta G_C(x,x')&=&\frac{1}{4\pi} \int_0^{\infty} d\w_1 \int_0^{\infty} d\w_2 \frac{1}{\sqrt{\w_1 \w_2}}\Big\{[A_{\w \w_1}e^{-i\w_1 u}+B_{\w \w_1}e^{i\w_1 u}]\nonumber \\ &&
\times [A_{\w \w_2}^{*}e^{i\w_2 u'}+B_{\w \w_2}^{*}e^{-i\w_2 u'}]+[A_{\w \w_1}^{*}e^{i\w_1 u}+B_{\w \w_1}^{*}e^{-i\w_1 u}]\nonumber \\ &&
 \times [A_{\w \w_2}e^{-i\w_2 u'}+B_{\w \w_2}e^{i\w_2 u'}]\nonumber \\ &&
-\mbox{subtraction terms}\Big\}\;,
\eea
where the subtraction terms have exactly the same form except that the matching coefficients are replaced by the Bogolubov coefficients~\eqref{alphaK-betaK} for the Unruh state.
Then
\bea
\Big[\Delta G_C(x,x')\Big]_{;r;t'}&=&\frac{1}{4\pi}\frac{1}{(1-\frac{2M}{r})} \int_0^{\infty} d\w_1 \int_0^{\infty} d\w_2 \sqrt{\w_1 \w_2}\Big\{[iA_{\w \w_1}e^{-i\w_1 u}-iB_{\w \w_1}e^{i\w_1 u}]\nonumber \\ &&
.[iA_{\w \w_2}^{*}e^{i\w_2 u'}-iB_{\w \w_2}^{*}e^{-i\w_2 u'}]+[-iA_{\w \w_1}^{*}e^{i\w_1 u}+iB_{\w \w_1}^{*}e^{-i\w_1 u}]\nonumber \\ &&
.[-iA_{\w \w_2}e^{-i\w_2 u'}+iB_{\w \w_2}e^{i\w_2 u'}]\nonumber \\ &&
-\mbox{subtraction terms}\Big\}\;. \label{Delta-GCrtp}
\eea
A similar calculation for $\Big[\Delta G_C(x,x')\Big]_{;t;r'}$ gives the opposite sign for each term in square brackets and a replacement of $r$ with $r'$ in the overall factor of $(1-\frac{2M}{r})^{-1}$.
Thus
\be \lim_{x\to x'} \Big[\Delta G_C(x,x')\Big]_{;r;t'} = \lim_{x\to x'} \Big[\Delta G_C(x,x')\Big]_{;t;r'} \;. \label{rtptrp} \ee

Next consider
\bea
\Big[\Delta G_C(x,x')\Big]_{;t;t'}&=&\frac{1}{4\pi} \int_0^{\infty} d\w_1 \int_0^{\infty} d\w_2 \sqrt{\w_1 \w_2}\Big\{[-iA_{\w \w_1}e^{-i\w_1 u}+iB_{\w \w_1}e^{i\w_1 u}]\nonumber \\ &&
.[iA_{\w \w_2}^{*}e^{i\w_2 u'}-iB_{\w \w_2}^{*}e^{-i\w_2 u'}]+[iA_{\w \w_1}^{*}e^{i\w_1 u}-iB_{\w \w_1}^{*}e^{-i\w_1 u}]\nonumber \\ &&
.[-iA_{\w \w_2}e^{-i\w_2 u'}+iB_{\w \w_2}e^{i\w_2 u'}]\nonumber \\ &&
-\mbox{subtraction terms}\Big\}\;.  \label{Delta-GCttp}
\eea
A similar computation for $\Big[\Delta G_C(x,x')\Big]_{;r;r'}$ gives the relation
\be \Big[\Delta G_C(x,x')\Big]_{;r;r'} = \frac{1}{(1-\frac{2M}{r}) (1-\frac{2M}{r'})} \Big[\Delta G_C(x,x')\Big]_{;t;t'} \;. \label{rrpttp} \ee
Also a comparison of~\eqref{Delta-GCrtp} and~\eqref{Delta-GCttp} shows  that
\bea
\Big[\Delta G_C(x,x')\Big]_{;r;t'}=-\frac{\Big[\Delta G_C(x,x')\Big]_{;t;t'}}{1-\frac{2M}{r}}\;.
\label{rtpttp}
\eea
Finally by substituting~\eqref{rrpttp} into~\eqref{T_ttB} and substituting~\eqref{rtptrp} and~\eqref{rtpttp} into~\eqref{T_trB} one can see that
\bea
  \Delta \langle  T_{rt}\rangle = - \frac{\Delta \langle T_{tt}\rangle}{1-\frac{2M}{r}}\;. \label{DTrt-DTtt}
\eea

\end{document}